\documentclass[12pt,a4paper]{article}
\pdfoutput=1

\usepackage{geometry}
\usepackage{amsmath}
\usepackage{amssymb}
\usepackage[dvips]{graphicx}
\usepackage{cite}
\usepackage{pstricks}
\usepackage{bm}
\usepackage{pbox}
\usepackage{placeins}
\usepackage{graphicx}
\usepackage{caption}
\usepackage{subcaption}
\usepackage[T1]{fontenc}
\usepackage{footnote}
\usepackage{pdfpages}
\usepackage{hhline}
\usepackage{multirow}
\usepackage{enumitem}
\usepackage[toc,page]{appendix}
\usepackage{dsfont}
\allowdisplaybreaks

\definecolor{MyDarkBlue}{rgb}{0.1, 0.1, 0.8} %defining the color 'MyDarkBlue'
\definecolor{MyLightBlue}{rgb}{0.22,0.51,0.9}
\definecolor{MyGreen}{rgb}{0.0, 0.5, 0.0}
\definecolor{BrickRed}{rgb}{0.8, 0.25, 0.33}
\usepackage[colorlinks=true,linkcolor=blue,citecolor=MyDarkBlue,
urlcolor=MyLightBlue,bookmarksnumbered=true,bookmarksopen]{hyperref}
\hypersetup{colorlinks, citecolor=BrickRed,linkcolor=MyDarkBlue, urlcolor=MyLightBlue}

\begin{document}
\vspace*{-0.2in}
\begin{flushright}
OSU-HEP-19-02
\end{flushright}
\vspace{0.5cm}
\begin{center}
{\Large\bf On the Origin of Two-Loop Neutrino Mass 
\\from SU(5) Grand Unification      }\\
\end{center}

\vspace{0.5cm}
\renewcommand{\thefootnote}{\fnsymbol{footnote}}
\begin{center}
{\large
{}~\textbf{Shaikh Saad}\footnote{ E-mail: \textcolor{MyLightBlue}{shaikh.saad@okstate.edu}}
}
\vspace{0.5cm}

{\em Department of Physics, Oklahoma State University, Stillwater, OK 74078, USA }
\end{center}

%\vspace{0.6cm}
\renewcommand{\thefootnote}{\arabic{footnote}}
\setcounter{footnote}{0}
\thispagestyle{empty}

%%%%%%%%%%%%%%%%%%%%%%%%%%%%%%%%%%%%%%%%%%%%%%%
%%%%%%%%%%%%%%%%%%%%%%%%%%%%%%%%%%%%%%%%%%%%%%%
\begin{abstract}
In this work we propose  a  renormalizable model based on the $SU(5)$ gauge  group  
where neutrino mass originates at the two-loop level without extending the fermionic content of the  Standard Model (SM).
Unlike the conventional $SU(5)$ models, in this proposed scenario, neutrino mass is intertwined with the charged fermion masses. In addition to correctly reproducing the  SM charged fermion masses and mixings, neutrino mass  is generated at the quantum level, hence naturally explains the smallness  of neutrino masses.
In this set-up, we provide examples of gauge coupling unification that simultaneously  satisfy the proton decay constraints.  This model has the potential to be tested experimentally by measuring the proton decay in the future experiments.  Scalar leptoquarks that are naturally contained within this framework can accommodate the recent B-physics anomalies. 
\end{abstract}

%\tableofcontents
\newpage
\setcounter{footnote}{0}
%%%%%%%%%%%%%%%%%%%%%%%%%%%%%%%%%%%%%%%%%%%%%%%
%%%%%%%%%%%%%%%%%%%%%%%%%%%%%%%%%%%%%%%%%%%%%%%
\section{Introduction}\label{SEC-01}
Even though the Standard Model (SM) of particle physics is highly successful in describing the interactions of the fundamental particles, it has several drawbacks, such as not being able to explain the neutrino mass, existence of Dark Matter (DM) and the origin of matter-antimatter asymmetry of the universe. Among them, one of the most important downside of the SM is that the neutrinos remain  massless. However, experimentally neutrino oscillation has been observed, hence the neutrinos must have acquired mass in some unspecified mechanism yet to be discovered. Due to these short coming, the SM begs for extensions. Grand Unified Theories (GUTs) \cite{Pati:1974yy,Georgi:1974sy,Georgi:1974yf} are the leading candidates beyond the Standard Model (BSM) since they are ultraviolet complete theories and come with many aesthetic features. 
Among different possibilities, $\bm{SU(5)}$ GUT  is the simplest choice, this is the only simple group that contains SM gauge group as a subgroup and has the same rank as the SM gauge group. $\bm{SU(5)}$ GUT can incorporate gauge coupling unification, it relates quarks with leptons and  quantization of the electric charge can also be understood. 

In the first proposed $\bm{SU(5)}$ GUT by Georgi and Glashow \cite{Georgi:1974sy}, 
the three families of fermions of the SM belong to the ${\bm{\overline{5}_F}}_i+{\bm{10_F}}_i$ ($i=1-3$ is the generation index) representations of the $\bm{SU(5)}$. Quarks and leptons are unified in these representations as
can be seen from their decompositions  under the SM:
\begin{align}
&{\bm{\overline{5}_F}}_i= \ell_i (1,2,-\frac{1}{2}) \oplus d_i^c (\overline{3},1,\frac{1}{3}),  \label{GG1}\\
&{\bm{10_F}}_i= q_i (3,2,\frac{1}{6}) \oplus u_i^c (\overline{3},1,-2/3) \oplus e_i^c (1,1,1), \label{GG2} 
\end{align} 

\noindent
where, $\ell_i=(\nu_i\;\; e_i)^T$ and $q_i= (u_i \;\;d_i)^T$. Interestingly, these multiplets contain only the fermions that are present in the SM, no additional fermions need to be introduced to cancel the gauge anomalies.

To describe our universe, the $\bm{SU(5)}$ gauge symmetry needs to be broken to that of the SM at the high scale: $SU(5)\rightarrow SU(3)_C\times SU(2)_L\times U(1)_Y$ that  can be achieved by employing a Higgs field in the $\bm{24_H}$-dimensional
\footnote{ 
Alternatively, instead of $\bm{24_H}$ Higgs, $\bm{SU(5)}$ breaking to the SM group can also be achieved by using $\bm{75_H}$ Higgs \cite{Hubsch:1984pg,Hubsch:1984qi}.
}
representation \cite{Georgi:1974sy}. When this field acquires a vacuum expectation value (VEV) in the SM singlet direction, the GUT symmetry is spontaneously broken down to  the SM. Then at the low energy scale, the SM gauge symmetry is spontaneously broken by the $\bm{5_H}$-dimensional representation: $SU(3)_C\times SU(2)_L\times U(1)_Y \rightarrow SU(3)_C\times U(1)_{em}$. As a result, the SM Higgs contained in the $\bm{5_H}$-Higgs generates masses to all the charged fermions.   However, this scenario predicts special mass relations $m_e=m_d, m_{\mu}=m_s$ \cite{Georgi:1979df} at the GUT scale that is ruled out by the experimental data. The shortcomings of the Georgi-Glashow model are:
\begin{itemize}
\item[--] it predicts wrong mass relations among the charged fermions.
\item[--] it fails to achieve gauge coupling unification.
\item[--] neutrinos remain massless.
\end{itemize}

In the literature, several different attempts are made to solve the aforementioned problems of the  Georgi-Glashow model. In all these works, neutrinos receive mass either at the tree-level or at the one-loop level by extending the scalar and/or fermion sectors. Here on the contrary, we construct a viable model where neutrino mass appears at the two-loop level  without introducing any new fermions to the SM. So in our model, neutrinos are Majorana in nature.
Realistic charged fermion masses are generated at the tree-level and the neutrino masses are originated due to the quantum corrections, hence  naturally explains the lightness of the neutrino masses. We also show that the neutrino mass in this set-up does not decouple  but gets entangled with the charged fermion masses. We construct the scalar potential and compute the Higgs spectrum that are relevant for the study of the gauge coupling unification. Successful scenarios of gauge coupling unification are presented by properly taking into account the proton decay constraints. Proton decay rate in this scenario is expected to be within the experimental observable range. 
The novelty of this work is, 
 our proposed set-up is the first  construction of a renormalizable model based on $\bm{SU(5)}$ gauge symmetry without imposing any additional symmetries and without introducing any additional fermions  where the  neutrinos receive mass at the two-loop level.  We compare our proposed model with the existing realistic $\bm{SU(5)}$ GUT models in the literature in great details. Scalar leptoquakrs naturally contained within the representations  required to generate  realistic charged fermion and neutrino masses in our framework  can accommodate the recent B-physics  anomalies.

%%%%%%%%%%%%%%%%%%%%%%%%%%%%%%%%%%%%%%%%%%%%%%%
%%%%%%%%%%%%%%%%%%%%%%%%%%%%%%%%%%%%%%%%%%%%%%%
\section{Neutrino mass in renormalizable \texorpdfstring{$SU(5)$}{TEXT} GUTs }\label{SEC-02}
The first shortcoming of the Georgi-Glashow model listed above can be fixed in  two different ways: one approach to correct the bad mass relations is to add higher-dimensional operators \cite{Ellis:1979fg}, which we do not pursue.  The alternative approach that we are interested in, is to work within the renormalizable framework that requires extension of the minimal Higgs sector.  As aforementioned, the SM fermions belong to the ${\bm{\overline{5}_F}}+{\bm{10_F}}$-multiplets of $\bm{SU(5)}$:
\begin{align}
\bm{\overline{5}_F}=\begin{pmatrix}
d^c_1\\d^c_2\\d^c_3\\e\\ -\nu
\end{pmatrix},\;\;\;\bm{10_F}=\frac{1}{\sqrt{2}} \begin{pmatrix}
0&u^c_3&-u^c_2&u_1&d_1\\
-u^c_2&0&u^c_1&u_2&d_2\\
u^c_2&-u^c_1&0&u_3&d_3\\
-u_1&-u_2&-u_3&0&e^c\\
-d_1&-d_2&-d_3&-e^c&0
\end{pmatrix}.
\end{align}

\noindent
The Higgs fields that can generate masses for the charged fermions can be identified from the fermion bilinears \cite{Slansky:1981yr}: 
\begin{align}
&\bm{\overline{5}}\times \bm{10}= \bm{5}+\bm{45}, \label{bili1}\\
&\bm{10}\times \bm{10}= \bm{\overline{5}}_s+\bm{\overline{45}}_a+\bm{\overline{50}}_s, \label{bili2}
\end{align}

\noindent 
where the subscripts `s' and `a' represent symmetric and antisymmetric combinations. So the possible set of Higgs fields that can have Yukawa couplings is $\{ \bm{5_H}, \bm{45_H}, \bm{50_H} \}$. However, among them, only $\bm{5_H}$ and  $\bm{45_H}$ contain SM like Higgs that can break the EW symmetry and generate charged fermion masses. Hence, the only possibility is to add a $\bm{45_H}$ \cite{Georgi:1979df}  in the Georgi-Glashow model to correct the bad mass relations.  With this addition, the Yukawa part of the Lagrangian is given by \cite{Dorsner:2006dj}:
\begin{eqnarray}\label{YUK}
\mathcal{L}_{Y} &=& Y_{1,ij} {\bm{\overline{5}_F}}_i^{\alpha} \ {\bm{10_F}}_{\alpha \beta,j} \
 \ {\bm{5}^*_H}^{\beta} \ + \ Y_{2,ij}
 \ {\bm{\overline{5}_F}}^{\delta}_i \ {\bm{10_F}}_{\alpha \beta,j}
\ {\bm{45}^*_H}^{\alpha \beta}_{\delta} \ + \nonumber\\
&+&\epsilon^{\alpha \beta \gamma \delta r} \left( Y_{3,ij} \
{\bm{10_F}}_{\alpha \beta,i} \ {\bm{10_F}}_{\gamma \delta,j} \ {\bm{5_{H}}}_r \
+ \ Y_{4,ij} \ {\bm{10_F}}_{\alpha \beta,i} {\bm{10_F}}_{m \gamma,j}
{\bm{45_H}}_{\delta r}^m \right),
\end{eqnarray}

\noindent where $\bm{SU(5)}$ group indices are explicitly shown and  $i,j=1-3$ are the generation indices. From this Lagrangian, the down-type quark and the charged-lepton mass matrices are given by:
\begin{eqnarray}
M_D &=& Y_1 \ v_5^* \ +  \ Y_2 \ v_{45}^* ,\label{MD}\\
M_E &=& Y^T_1 \ v_5^* \ - 3 \ Y^T_2 \ v_{45}^* .\label{ME}
\end{eqnarray}

\noindent Here we have defined $v_5=\langle \phi^0_1 \rangle /(\sqrt{2})$ and $v_{45}=\langle \Sigma^0_1 \rangle /(-2 \sqrt{3})$, where the SM like weak doublets from $\bm{5_H}$ and $\bm{45_H}$ are identified as $\phi_1=\left( \phi^+_1\;\;\phi_1^0  \right)^T$ and $\Sigma_1=\left( \Sigma^+_1\;\;\Sigma_1^0  \right)^T$. This normalization follows the relation: $2v^2_5+12v^2_{45}=v^2$, with $v=174$ GeV. The above relations clearly violate the simple mass relations of the Georgi-Glashow model and,  in the
Yukawa sector, there are enough parameters  to fit all the charged fermions masses and mixings. In Eq. \eqref{YUK}, $Y_1, Y_2, Y_3, Y_4$ are arbitrary $3\times 3$ Yukawa matrices, and the up-quark mass matrix is not related to the down-quark and chagred lepton mass matrices and is a symmetric complex matrix, $M_U=M^T_U$. In this renormalizable model, one can also achieve gauge coupling unification and the model is safe from too rapid proton decay \cite{Dorsner:2006dj}. In Sec. \ref{SEC-05}, we reproduce the result of Ref. \cite{Dorsner:2006dj}  and present a plot to demonstrate successful  gauge coupling unification in this scenario.  So the minimal model extended by $\bm{45_H}$ Higgs can simultaneously solve the first two shortcomings of the Georgi-Glashow model listed above except the last one. This renormalizable  model consists of fermion fields given in Eqs. \eqref{GG1}-\eqref{GG2} and scalar fields as given below in Eqs. \eqref{GG3}-\eqref{GG5},  and for brevity we refer to this model as MRSU5 (minimal renormalizable $\bm{SU(5)}$ GUT) for the rest of the text.
\begin{align}
\bm{5_H} \equiv \phi = &\phi_1 (1,2,\frac{1}{2}) \oplus \phi_2 (3,1,-\frac{1}{3}),  \label{GG3}\\
\bm{24_H} \equiv \Phi  = &\Phi_1 (1, 1, 0) \oplus \Phi_2 (1, 3, 0)\oplus \Phi_3 (8, 1, 0) \oplus \Phi_4 (3, 2, -\frac{5}{6} ) \oplus \Phi_5 (3, 2, +\frac{5}{6} ), \label{GG4} \\
\bm{45_H} \equiv \Sigma = &\Sigma_1 (1,2,\frac{1}{2}) \oplus \Sigma_2 (3,1,-\frac{1}{3}) \oplus \Sigma_3 (\overline{3},1,\frac{4}{3}) \oplus \Sigma_4 (\overline{3},2,-\frac{7}{6}) \oplus \Sigma_5 (3,3-\frac{1}{3}) 
\nonumber \\ & 
\oplus \Sigma_6 (\overline{6},1,-\frac{1}{3}) \oplus \Sigma_7 (8,2,\frac{1}{2}).  \label{GG5}
\end{align}   

Note however that in this minimal model the neutrinos are  still  massless just like the SM. Extension must be made to make this set-up  phenomenologically viable. Here we briefly review the different  possibilities of generating non-zero neutrino mass
\footnote{For a recent general review on neutrino mass generation mechanisms  for Majorana type neutrinos see Ref.  \cite{Cai:2017jrq}.}
 in the context of renormalizable $\bm{SU(5)}$ GUT.  

\begin{itemize}
\item Tree-level:

To incorporate neutrino mass, the simplest possibility is to add at least two right-handed Majorana neutrinos $\nu^c$(1,1,0) to MRSU5 model that are   singlets of $\bm{SU(5)}$. This possibility  can give rise to neutrino masses by using the type-I seesaw mechanism \cite{ss1,ss2,ss3,ss4}. Since this extension involves GUT group singlets, this approach may not be aesthetic and it is preferable to have multiplets that are non-singlets under the gauge group. Neutrino mass can be generated via type-II seesaw scenario \cite{Magg:1980ut,Schechter:1980gr,Lazarides:1980nt,Mohapatra:1980yp}
  if a Higgs in the $\bm{15_H}$-dimensional representation\footnote{Extension by $\bm{15_H}$ Higgs was first considered  within the non-renormalizable $\bm{SU(5)}$ context  \cite{Dorsner:2005fq,Dorsner:2005ii,Dorsner:2006hw}.} is added to the MRSU5 \cite{Dorsner:2007fy}, it is because the $\bm{15_H}$-Higgs contains a iso-spin triplet $(1,3,1)\subset \bm{15_H}$. Another possibility is to add at least two copies of fermion multiplets in the adjoint $\bm{24_F}$-dimensional  representation\footnote{Extension by $\bm{24_F}$ fermions was first considered  within the non-renormalizable $\bm{SU(5)}$ context \cite{Bajc:2006ia}.}  to the MRSU5 \cite{Perez:2007rm},  this scenario generates neutrino mass in a combination of   type-III \cite{Foot:1988aq} and type-I seesaw mechanisms. This scenario makes use of the fermionic weak triplet lies in the adjoint representation, $(1,3,0)\subset \bm{24_F}$.  These are the simple possibilities to incorporate neutrino mass at the tree-level by extending the MRSU5 model by one (type-II) or more (type-I and type-III) multiplets.

\FloatBarrier
\begin{figure}[th!]
\centering
\includegraphics[scale=0.33]{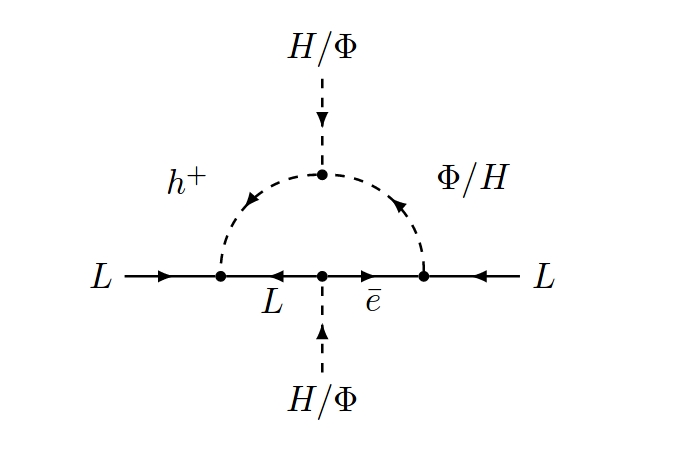}
\includegraphics[scale=0.3]{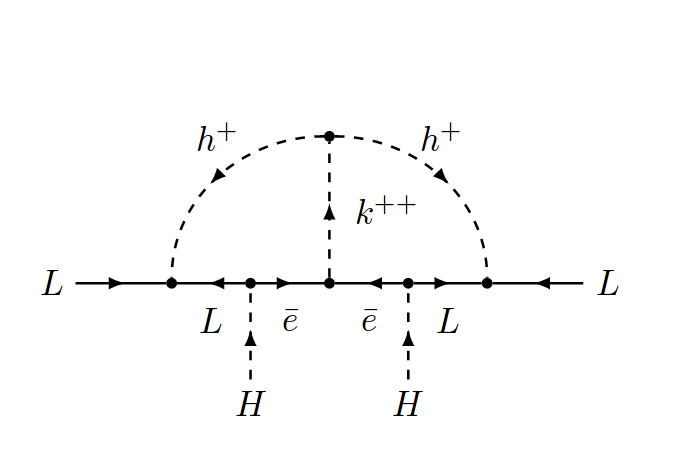}
\caption{ Left: Zee mechanism to generate neutrino mass at one-loop level. Right: Zee-Babu mechanism to generate neutrino mass at two-loop level.  }\label{zee-babu}
\end{figure}

\newpage
\item Loop-level: 

Neutrino mass can also be generated at the quantum level within the $\bm{SU(5)}$ GUT framework and is a very interesting alternative possibility. The first radiative model \footnote{First radiative model was proposed  for Dirac type neutrinos \cite{Cheng:1977ir}. Without introducing exotic fermions in the SM, generating Dirac mass for neutrinos are studied in great details recently in Ref. \cite{Saad:2019bqf}.}  of neutrino mass generation for Majorana type particles was proposed by Zee \cite{Zee:1980ai} in the context of the SM gauge group by  extending the SM  by another Higgs doublet and a singly charged scalar singlet. In Zee model, neutrino mass is generated at the one loop level as shown in Fig. \ref{zee-babu} (Feynman diagram on the  left). The $\bm{SU(5)}$ GUT embedding of Zee mechanism with the use of $\bm{10_H}$-dimensional Higgs was first proposed in Ref. \cite{Wolfenstein:1980sf}, also pointed out  in Ref. \cite{Barbieri:1981yw} and just recently studied in the renormalizable context in Ref.  \cite{Perez:2016qbo}. In this realization,   
the two SM like Higgs doublets are contained in 
the $\bm{5_H}$ and $\bm{45_H}$   and  the singly charged Higgs $h^+$(1,1,1) which is singlet under the SM lies in $\bm{10_H}$. Many different variations of the original  Zee model by extending the SM are  proposed in the literature. A particular model proposed in Ref. \cite{Brdar:2013iea} uses a real scalar triplet instead of the second Higgs doublet. In addition to the singly charged scalar, this model needs three copies of vector-like lepton doublets to incorporate  Zee mechanism. Just recently this one-loop Zee-type model is also embedded in a renormalizable $\bm{SU(5)}$ GUT   \cite{Kumericki:2017sfc} by extending the MRSU5 by both scalar and fermion multiplets. In their work, the real scalar triplet is embedded in $\bm{24_H}$, the singly charged scalar in $\bm{10_H}$  and the  vector-like lepton doublets in three generations of ${\bm{5_F}}_i+{\bm{\overline{5}_F}}_i$ matter fields. Hence, the  Georgi-Glashow model is extended by  $\bm{10_H}$ Higgs and three generations of vector-like leptons $\bm{5_F}+\bm{\overline{5}_F}$.  In the models mentioned above, the particle running in the loop are colorless, however, neutrino mass can also be generated via Zee mechanism while colored particles run through the loop. For such leptoquark mechanism of neutrino masses at the one-loop level within the $\bm{SU(5)}$ GUT framework, see for example Ref. \cite{Dorsner:2017wwn}. 

\end{itemize}

%%%%%%%%%%%%%%%%%%%%%%%%%%%%%%%%%%%%%%%%%%%%%%%
%%%%%%%%%%%%%%%%%%%%%%%%%%%%%%%%%%%%%%%%%%%%%%%
\section{The proposed model}\label{SEC-03}
In the models discussed above, the 
neutrino mass can be generated at the tree-level via seesaw mechanism or at the one-loop level via Zee mechanism within the context of renormalizable  $\bm{SU(5)}$ GUT.  In this work, we for the first time construct a realistic  $\bm{SU(5)}$ GUT, where the origin of the neutrino mass is realized at the two-loop level. In our construction, we restrict ourselves by demanding the following requirements:
\begin{itemize}
\item[--] model should be renormalizable.
\item[--] the only symmetry of the theory is 
$\bm{SU(5)}$ gauge symmetry.
\item[--] no additional fermions are added compared to the already existing ones in the SM.
\item[--] automatic vanishing of the neutrino mass at the tree-level and at the one-loop level.
\item[--] no $\bm{SU(5)}$ singlet is allowed.
\item[--] in our search, we restrict ourselves to the representations  of dimension $< 100$. 
\end{itemize}

As aforementioned, the Higgs fields that can have Yukawa interactions can be determined from the fermion bilinears presented in Eqs. \eqref{bili1}-\eqref{bili2}. In search of the simplest renormalizable $\bm{SU(5)}$ GUT where two-loop neutrino mass mechanism can be realized, we take a closer look at these fermion bilinears.     As noted above, out of $\{ \bm{5_H}, \bm{45_H}, \bm{50_H} \}$ Higgs multiplets,  $\bm{5_H}$ and $\bm{45_H}$-multiplets contain the SM like doublets, hence contribute to the generation of charged fermion masses. Due to the absence of right-handed neutrino, no such Dirac mass term is allowed for the neutrinos.  However, in this work, we show that the presence of the Yukawa coupling  of the $\bm{50}_H$ Higgs to the fermions
 in Eq. \eqref{bili2}  plays an important role in generating non-zero neutrino mass via two-loop mechanism.  We show that this new Yukawa interaction:
\begin{align}\label{YUK50}
\mathcal{L}_Y\supset {Y_5}_{ij} {\bm{10_F}}_{\alpha\beta,i} {\bm{10_F}}_{\gamma\delta,j} {\bm{50_H}}^{\alpha\beta\gamma\delta},
\end{align}

\noindent
along with the already existing Yukawa interactions given in Eq. \eqref{YUK} combinedly   determine the neutrino mass. Hence, neutrino mass does not appear to be completely detached, rather gets intertwined with the charged fermion masses. Here the Yukawa coupling $Y_5$ is a symmetric $3\times 3$ matrix in the generation space.  Since neutrino mass appears at the two-loop level in this model, the neutrino masses are highly suppressed compared to the charged fermions, hence naturally explains the smallness of the neutrino masses.  The decomposition of this Higgs fields under the SM is as follows: 
\begin{align}\label{Higgs50}
&\bm{50_H} \equiv \chi = \chi_1 (1,1,-2) \oplus \chi_2 (3,1,-\frac{1}{3}) \oplus \chi_3 (\overline{3},2,-\frac{7}{6}) \oplus \chi_4 (6,1,\frac{4}{3}) \oplus \chi_5 (\overline{6},3,-\frac{1}{3}) \oplus \chi_6 (8,2,\frac{1}{2}). 
\end{align}

In the context of the SM gauge group, the possibility of generating neutrino mass via two-loop
 is well known   \cite{Cheng:1980qt}. The simplest possibility is to add a singly charged scalar and a doubly charged scalar both singlets under the SM group and is commonly known as Zee-Babu model \cite{Babu:1988ki} as shown in Fig. \ref{zee-babu} (Feynman diagram on the right).   Many variations of the Zee-Babu model are proposed in the literature by extending the SM particle content. Note that in both the one-loop (Zee model) and the two-loop (Zee-Babu model) neutrino mass mechanism, at least  two new multiplets need to be added to the theory to generate non-zero neutrino mass. In the original Zee model, in addition to a second SM like Higgs doublet, a singly charged scalar singlet needs to be added. In the original version of the  Zee-Babu model, again two BSM multiplets, one singly charged singlet and one doubly charged singlet need to be introduced.    Below, we show that to realize two-loop neutrino mass in the context of renormalizable $\bm{SU(5)}$ GUT, at least two new multiplets need to be added to the MRSU5.

Our framework  incorporates  Zee-Babu mechanism  to explain the extremely small neutrino mass. The Yukawa coupling given in Eq. \eqref{YUK50} has doubly charged scalar couplings to two charged leptons (suppressing the group indices): 
\begin{align}
{Y_5}_{ij} {\bm{10_F}}_i{\bm{10_F}}_j\bm{50_H}\supset {Y_5}_{ij}\; \ell_i^c\; \ell_j^c\; \chi_1^{--},\;\;\;(\ell_i^c= e^c, \mu^c, \tau^c).
\end{align}

\noindent
Now to complete the loop-diagram,  one must introduce at least one more Higgs multiplet, which is however not arbitrary but unambiguously determined   by the group theory. The simplest possibility is to add a  $\bm{40_H}$-dimensional representation that has the following decomposition under the SM: 
\begin{align}\label{Higgs40}
&\bm{40_H} \equiv \eta = \eta_1 (1,2,-\frac{3}{2}) \oplus \eta_2 (\overline{3},1,-\frac{2}{3}) \oplus \eta_3 (3,2,\frac{1}{6}) \oplus \eta_4 (\overline{3},3,-\frac{2}{3}) \oplus \eta_5 (\overline{6},2,\frac{1}{6}) \oplus \eta_6 (8,1,1).
\end{align}

\noindent
Note that $\bm{40_H}$ Higgs has an iso-spin doublet $\eta_1=(\eta_1^-\; \eta_1^{--})^T$ with a hypercharge of $Y=-3/2$ that is necessary to close the loop-diagram. The $\bm{SU(5)}$ invariant scalar potential contains cubic terms relevant for neutrino mass generation that are of the form:  
\begin{align}\label{CUBIC}
V &\supset \mu_1\; {\bm{5_H}}_{\gamma}{\bm{50_H}}^{\alpha\beta\gamma\delta}{{\bm{40_H}}^{\ast}}_{\alpha \beta \gamma} + \mu_2\; {\bm{45_H}}^{\rho}_{\gamma\delta}{\bm{50_H}}^{\alpha\beta\gamma\delta}{{\bm{40_H}}^{\ast}}_{\alpha \beta \rho}
\nonumber \\ &
\supset \chi_1^{--}\; \eta^{+}_1 \left( \left( \frac{1}{\sqrt{2}}\right) \mu_1 \; \phi_1^+ + \left(-\frac{\sqrt{3}}{2}\right) \mu_2 \; \Sigma_1^+\; \right)
\nonumber \\ &+
 \chi_1^{--}\; \eta^{++}_1 \left( \left( -\frac{1}{\sqrt{2}}\right) \mu_1 \;  \phi_1^0  + \left(\frac{\sqrt{3}}{2}\right) \mu_2 \;  \Sigma_1^0  \right)
. 
\end{align}

\noindent Where $\phi_1^+\subset \phi_1(1,2,\frac{1}{2})$ and $\Sigma_1^+\subset \Sigma_1(1,2,\frac{1}{2})$ are the singly charged scalars from the SM like doublets. And the  relevant quartic terms in the potential to complete the loop-diagram are of the form:
\begin{align}\label{QUARTIC}
V &\supset {\bm{40_H}}^{\alpha\beta\gamma} \left(
\lambda {\bm{5_H}}_{\gamma}{\bm{5_H}}_{\delta}{\bm{45_H}}^{\delta}_{\alpha\beta}
+
\lambda^{\prime} {\bm{5_H}}_{\sigma}{\bm{45_H}}^{\delta}_{\alpha\beta}{\bm{45_H}}^{\sigma}_{\gamma\delta}
\right)  
    \nonumber \\ & \supset
 \eta_1^{-} \phi_1^+ \Sigma_1^0  
 \left( 
 \left(\frac{\sqrt{3}}{2}\right) \lambda  \phi_1^0 
 +  \left(-\frac{3}{4\sqrt{2}}\right) \lambda^{\prime}  \Sigma_1^0 
    \right) 
\nonumber \\ &   
 +
 \eta_1^- \Sigma_1^+  \phi_1^0 
 \left(  
  \left(\frac{-\sqrt{3}}{2}\right) \lambda  \phi_1^0  
 +  \left(\frac{3}{4\sqrt{2}}\right) \lambda^{\prime} \Sigma_1^0
    \right)    
\nonumber \\ & + 
 \eta_1^{--} \phi_1^+ \Sigma_1^0  
 \left( 
 \left(\frac{\sqrt{3}}{2}\right) \lambda \phi_1^+ 
 +  \left(-\frac{3}{4\sqrt{2}}\right) \lambda^{\prime}  \Sigma_1^+
    \right) 
\nonumber \\ &
    +
 \eta_1^{--} \Sigma_1^+ \phi_1^0
 \left(  
  \left(\frac{-\sqrt{3}}{2}\right) \lambda \phi_1^+  
 +  \left(\frac{3}{4\sqrt{2}}\right) \lambda^{\prime}  \Sigma_1^+
    \right).    
\end{align}

\begin{figure}[th!]
\centering
\includegraphics[scale=0.45]{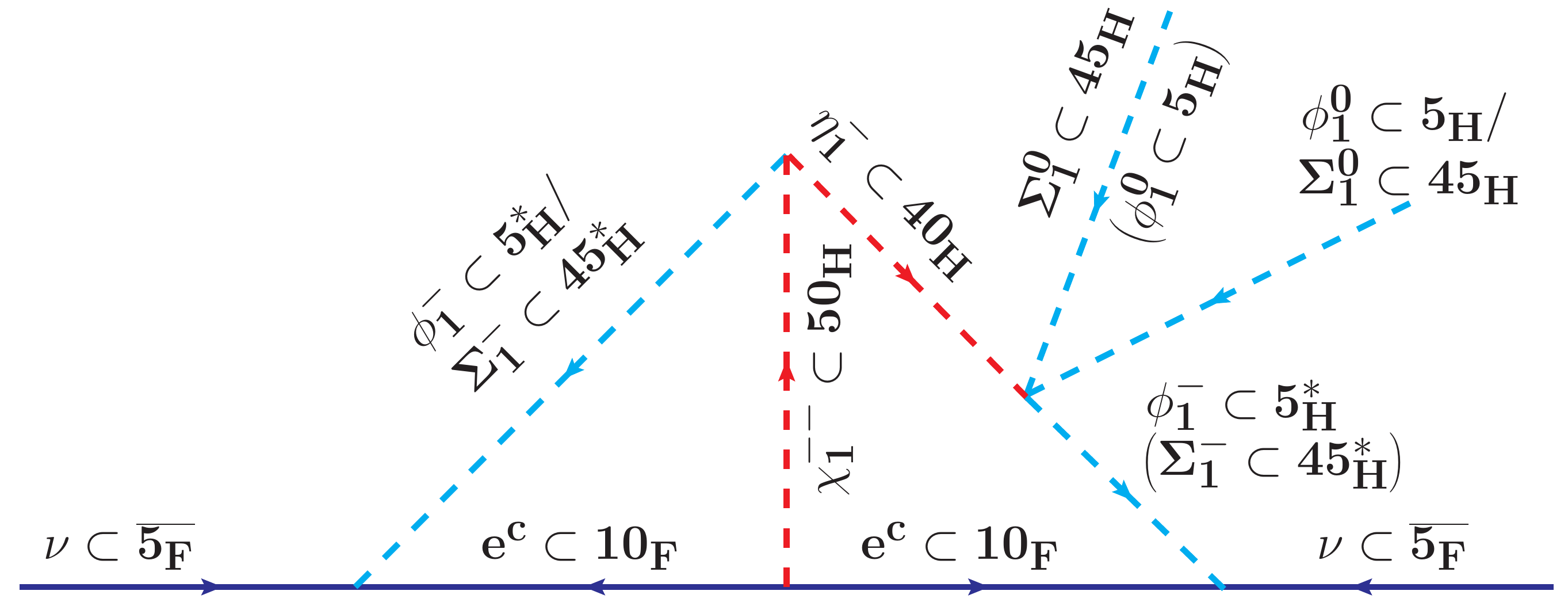}\\
\vspace{0.5in}
\includegraphics[scale=0.4]{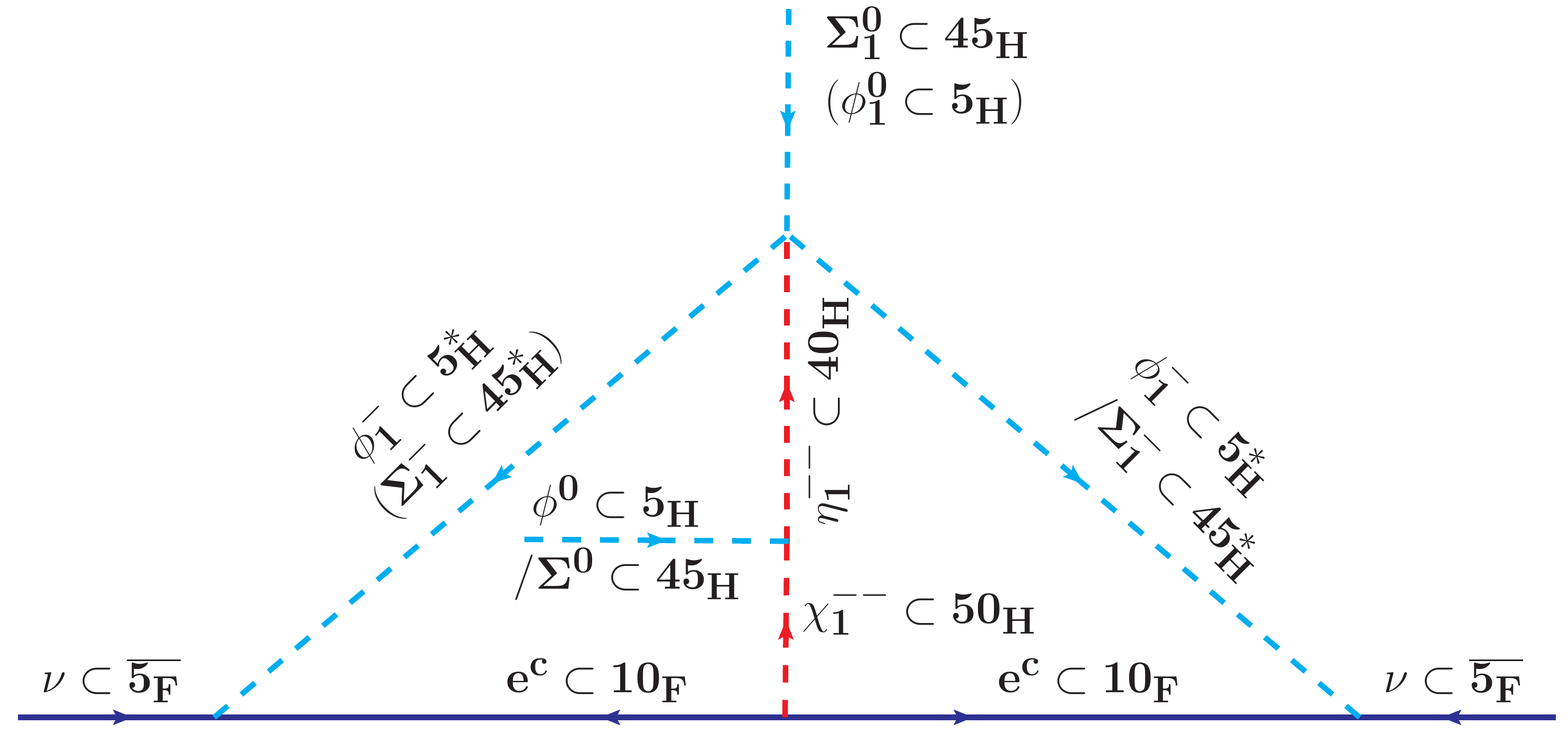}
\caption{Two-loop Feynman diagrams responsible for neutrino mass geenration in our proposed renormalizable $\bm{SU(5)}$ GUT. The propagators in red are the multiplets that belong to the $\bm{40_H}$ and $\bm{50_H}$ representations. For each of these diagrams, there is a second set of diagrams that can be achevied by replacing the  multiplet by the associated multiplet shown in the parenthesis.  }\label{babu-su5}
\end{figure}

With the simultaneous presence of the Yukawa coupling Eq. \eqref{YUK50}, the scalar cubic couplings Eq. \eqref{CUBIC} and the scalar quartic couplings Eq. \eqref{QUARTIC}, the accidental  global $U(1)_{B-L}$ is broken that is required to generate non-zero neutrino mass.   
With these relevant cubic and quartic terms in the scalar potential,  the diagrams 
 responsible for generating neutrino mass in our model
is presented are  Fig. \ref{babu-su5}. 
The scalar multiplets beyond the MRSU5 running in the loop belonging to the $\bm{40_H}$ and $\bm{50_H}$ representations are shown in red. These BSM particle contributing to the generation of neutrino mass are expected to live at scales much below the GUT scale.  
Note that similar diagrams with colored particles running in the loop can also be drawn. For example, instead of $(1,2,\frac{1}{2})\subset \bm{5_H}$ running in the loop, one can replace it by $(3,1,-\frac{1}{3})\subset \bm{5_H}$. However, since these colored triplets mediate dangerous   proton decay, their masses are assumed to be of the order of GUT scale to suppress the proton decay rate, hence we do not consider such diagrams, but can be trivially included.
In the context of the SM, 
similar  diagrams as shown in  Fig. \ref{babu-su5}  are realized recently by extending  the SM with three new fields, a doubly charged scalar singlet and two doublets, a second SM-like doublet with hypercharge of 1/2 and a third doublet with hypercharge 3/2 in two different works \cite{Law:2013dya, Cao:2017xgk}.

\FloatBarrier

Here we compute the neutrino mass matrix. First note that the 
 breaking of the EW symmetry allows mixings of the particles carrying the same electric  charge. Mixing among the singly charged fields are induced by the quartic terms of Eq. \eqref{QUARTIC}, whereas for the doubly charged particles are induced by the cubic terms
 of Eq. \eqref{CUBIC}. After the breaking of the EW symmetry,  the part of the  scalar potential  containing the relevant mixing terms are given by:
 
\newpage 
\begin{align}\label{mixing}
V\supset 
&
\begin{pmatrix}
\phi^0_1&\Sigma^0_1
\end{pmatrix}
\begin{pmatrix}
{m^0_{\phi}}^2&{m^0_{12}}^2\\
{m^0_{12}}^2&{m^0_{\Sigma}}^2
\end{pmatrix}
\begin{pmatrix}
\phi^{0\ast}_1\\ \Sigma^{0\ast}_1
\end{pmatrix}
+
\begin{pmatrix}
\chi^{--}_1&\eta^{--}_1
\end{pmatrix}
\begin{pmatrix}
{m^{++}_{\chi}}^2&{m^{++}_{12}}^2\\
{m^{++}_{12}}^2&{m^{++}_{\eta}}^2
\end{pmatrix}
\begin{pmatrix}
\chi^{++}_1\\ \eta^{++}_1
\end{pmatrix}
\nonumber \\ &+
\begin{pmatrix}
\phi^{+}_1&\Sigma^{+}_1&\eta^{+}_1
\end{pmatrix}
\begin{pmatrix}
{m^{+}_{\phi}}^2&{m^{+}_{12}}^2&{m^{+}_{13}}^2\\
{m^{+}_{12}}^2&{m^{+}_{\Sigma}}^2&{m^{+}_{23}}^2\\
{m^{+}_{13}}^2&{m^{+}_{23}}^2&{m^{+}_{\eta}}^2
\end{pmatrix}
\begin{pmatrix}
\phi^{-}_1\\ \Sigma^{-}_1\\ \eta^{-}_1
\end{pmatrix}.
\end{align}

\noindent 
Here the off-diagonal entries are the mixing terms as already mentioned above and the diagonal entries are the mass terms which for simplicity we do not write down explicitly, however can be computed straightforwardly from the full potential.  In the next section, we will construct part of the scalar potential that is relevant for the study of the gauge coupling unification.  
For simplicity, treating all the parameters of the the scalar potential appearing in Eq. \eqref{mixing} to be real, the transformation between the weak basis and the mass basis for the CP-even neutral fields, the doubly charged scalars and the singly charged scalars can be written as:
\begin{align}
&\begin{pmatrix}
\phi^{0}_1\\ \Sigma^{0}_1
\end{pmatrix}
=
\begin{pmatrix}
\cos\phi&\sin\phi\\
-\sin\phi&\cos\phi
\end{pmatrix}
\begin{pmatrix}
H^{0}_1\\ H^{0}_2
\end{pmatrix},
\\
&\begin{pmatrix}
\chi^{--}_1\\ \eta^{--}_1
\end{pmatrix}
=
\begin{pmatrix}
\cos\omega&\sin\omega\\
-\sin\omega&\cos\omega
\end{pmatrix}
\begin{pmatrix}
H^{--}_1\\ H^{--}_2
\end{pmatrix},
\\
&\begin{pmatrix}
\phi^{+}_1\\ \Sigma^{+}_1\\ \eta^{+}_1
\end{pmatrix}
=
\begin{pmatrix}
1&0&0\\
0&\cos\theta_{23}&\sin\theta_{23}\\
0&-\sin\theta_{23}&\cos\theta_{23}
\end{pmatrix}
\begin{pmatrix}
\cos\theta_{13}&0&\sin\theta_{13}\\
0&1&0\\
-\sin\theta_{13}&0&\cos\theta_{13}
\end{pmatrix}
\begin{pmatrix}
\cos\theta_{12}&\sin\theta_{12}&0\\
-\sin\theta_{12}&\cos\theta_{12}&0\\
0&0&1
\end{pmatrix}
\begin{pmatrix}
G^{+}\\ H^{+}_1\\ H^{+}_2
\end{pmatrix}.
\end{align} 
 
\noindent 
Here, the fields labeled  with $H$ ($H^+_i, H^{++}_i$) represent the mass eigenstates and $G^+$ is the Goldstone boson.  
This leads to:
\begin{align}
\mathcal{L} &\supset 
Y_1 \bm{\overline{5}_F} \bm{10_F} \bm{5^{\ast}_H} 
+
Y_1 \bm{\overline{5}_F} \bm{10_F} \bm{45^{\ast}_H} 
+
Y_5 \bm{10_F}\bm{10_F}\bm{50_H}
\\ &
\supset
{\nu_{L}}_i\ell^c_j \left( H^-_1 {Y^+_1}_{ij} + H^-_2 {Y^+_2}_{ij} \right)
+
\ell^c_i \ell^c_j\left(  H^{--}_1 {Y^{++}_1}_{ij} + H^{--}_2 {Y^{++}_2}_{ij} \right) .
\end{align}
 
\noindent 
Where we have defined,
\begin{align}
&Y^+_1=
\frac{1}{\sqrt{2}}Y_1 \left( c_{13}s_{12} \right)
+ 
\frac{\sqrt{3}}{2}Y_2 \left( c_{12}c_{23}-s_{12}s_{13}s_{23}  \right),
\\
&Y^+_2=
\frac{1}{\sqrt{2}}Y_1 \left( s_{13} \right)
+ 
\frac{\sqrt{3}}{2}Y_2 \left( c_{13}s_{23}  \right),
\\
&Y^{++}_1=
Y_5 c_{\omega},
\\&Y^{++}_2=
Y_5 s_{\omega},
\end{align} 
 
\noindent 
and we have made use of the  notation: $c_{\omega}=\cos\omega, s_{\omega}=\sin\omega, c_{ij}=\cos_{\theta_{ij}}, s_{ij}=\sin_{\theta_{ij}}$.

\noindent 
Furthermore,  we get:
\begin{align}
V &\supset \mu_1\; {\bm{5_H}}_{\gamma}{\bm{50_H}}^{\alpha\beta\gamma\delta}{{\bm{40_H}}^{\ast}}_{\alpha \beta \gamma} + \mu_2\; {\bm{45_H}}^{\rho}_{\gamma\delta}{\bm{50_H}}^{\alpha\beta\gamma\delta}{{\bm{40_H}}^{\ast}}_{\alpha \beta \rho}
\nonumber \\
&\supset 
\left(
\mu_{11} H^+_1H^+_1
+
\mu_{22} H^+_2H^+_2
+
\mu_{12}H^+_1H^+_2
+
\mu_{21} H^+_2H^+_1 \right) \left(c_{\omega}H^{--}_1+s_{\omega}H^{--}_2\right).
\end{align} 
 
\noindent 
Where,
\begin{align}
&\mu_{11}= \left( -c_{12}s_{23}-s_{12}c_{23}s_{13}\right) \widetilde{\mu}_1 
,\;\;
\mu_{22}= \left( c_{23}c_{13}\right) \widetilde{\mu}_2, 
\\
&\mu_{12}= \left( -c_{12}s_{23}-s_{12}c_{23}s_{13}\right) \widetilde{\mu}_2
,\;\;
\mu_{21}= \left( c_{23}c_{13}\right) \widetilde{\mu}_1, 
\\
&\widetilde{\mu}_1=
\frac{\mu_1}{\sqrt{2}}\left(c_{13}s_{12}\right)-\frac{\sqrt{3}\mu_2}{2} 
\left( c_{12}c_{23}-s_{12}s_{13}s_{23}\right),
\\
&\widetilde{\mu}_2=
\frac{\mu_1}{\sqrt{2}}\left(s_{13}\right)-\frac{\sqrt{3}\mu_2}{2} 
\left( c_{13}s_{23}\right).
\end{align}

Then, the neutrino mass matrix is evaluated to be:
\begin{align}
{\mathcal{M}_{\nu}}_{ij}
&=
\mu_{AB}
{Y^+_A}_{ik}{Y_5^{\ast}}_{kl}{Y^+_B}_{lj}
\left[ c^2_{\omega}I_{1ABkl}+s^2_{\omega}I_{2ABkl}  \right] + \text{transpose}.
\end{align} 
 
\noindent 
It contains four terms corresponding to $AB=\{11,22,12,21\}$. 
Here the sum over the repeated indices $k$ and $l$ is  understood.   
The loop function is $I_{cabkl}= 
I\left( m^+_a,m^+_b,m^{++}_c,m_k,m_l  \right)$, where $m_{k,l}$ are the mass of the SM fermions, $m^+$ and $m^{++}$ are the mass of the singly and the doubly charged scalars  running inside the loop. We evaluate this loop function   as follows. 
To make life simple, we assume   
$m^{+}_1=m^{+}_2$
and furthermore use the approximation $m_{c,a,b,}>>m_{k,l}$,  which is valid since charged lepton masses are small compared to the BSM charged scalars running in through the loop. Then one finds \cite{Cao:2017xgk},
\begin{align}\label{loop-function}
I_{cabkl}\approx I_{ca}=
\frac{1}{(16\pi^2)^2}
 \int_0^1dx\int_0^1dy
 \left[  \frac{-\text{ln}\left(\Delta_{ca}\right)}{1-\Delta_{ca}}  \right],
\end{align} 

\noindent
with,
\begin{align}
\Delta_{ca}=\frac{(1-y)r^2_{ca}+y(1-x)}{y(1-y)};\;\;\;
r_{ca}=\frac{m^{++}_c}{m^{+}_a}.
\end{align}

\FloatBarrier
\begin{figure}[th!]
\centering
\includegraphics[scale=0.4]{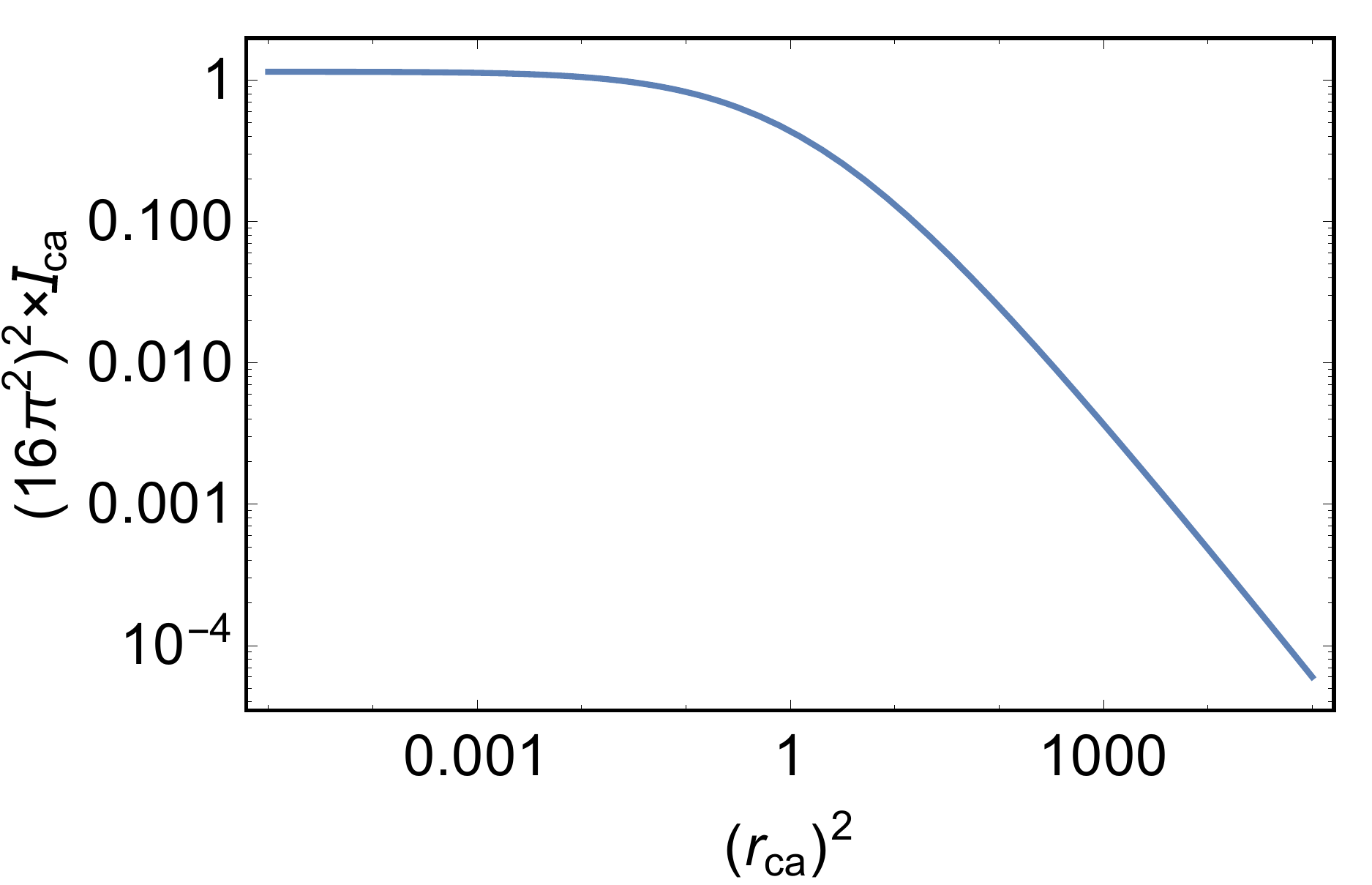}
\caption{The behaviour of the loop function given in Eq. \eqref{loop-function} as a function of $r_{ca}^2=\left( \frac{m^{++}_c}{m^{+}_a} \right)^2$. }\label{loop-function-figure}
\end{figure}

\noindent
Note that the loop function corresponding to our diagram is different from the one that appears in the conventional Zee-Babu model, it is due to different chirality structure. Unlike the conventional Zee-Babu model, chirality flip of the fermions do not take place inside the loop.  The behavior of the loop function as a function of $r^2_{ca}$ is presented in Fig. \ref{loop-function-figure}.

Here we note that the neutrino masses do not decouple from that of the charged fermion masses, rather gets entangled with them. To find the correlation,  we  express the Yukawa couplings $Y_1$ and $Y_2$ in terms of the down-quark and charged-lepton masses matrices from Eqs. \eqref{MD}-\eqref{ME} as:
\begin{align}
&Y_1=\frac{\sqrt{2}}{4v_1}\left( 3 M_D+M^T_E\right),\\
&Y_2=\frac{-\sqrt{3}}{2v_2}\left( M_D-M^T_E\right).
\end{align}  

\noindent 
As a result, $Y^+_{1,2}$ can also be expressed in terms of down-quark and charged-lepton masses matrices:
\begin{align}
&Y^+_1=
\frac{3}{4}
\left(
\frac{c_{13}s_{12}}{v_1}
-\frac{c_{12}c_{23}-s_{12}s_{13}s_{23}}{v_2}
\right)
D^cM_D^{diag}V^{\dagger}_{CKM}
+\frac{1}{4}
\left(
\frac{c_{13}s_{12}}{v_1}
+\frac{3(c_{12}c_{23}-s_{12}s_{13}s_{23})}{v_2}
\right)
M_E^{diag},
\\
&Y^+_2=
\frac{3}{4}
\left(
\frac{s_{13}}{v_1}
-\frac{c_{13}s_{23}}{v_2}
\right)
D^cM_D^{diag}V^{\dagger}_{CKM}
+\frac{1}{4}
\left(
\frac{s_{13}}{v_1}
-\frac{3c_{13}s_{23}}{v_2}
\right)
M_E^{diag}.
\end{align} 

\noindent
Here we have gone to a basis where the up-quark and charged lepton mass matrices are diagonal. 
In this rotated basis, the $Y_5$ matrix takes the form: $E^cY_5{E^c}^T$.
To get to these relations, we used the following convention for diagonalization of the charged fermion masses:  
\begin{align}
M_U=U^cM_U^{diag}U,\;\;M_D=D^cM_D^{diag}D,\;\;M_E=E^cM_E^{diag}E,\;\;\;\; (U^c=U^T).
\end{align}

Here we provide an example of how 
 correct order of neutrino mass can be achieved. The neutrino mass has the from, $m_{\nu}\sim y^3\mu I$. To reproduce the correct  tau mass, the biggest entry ($33$ entry) needs to be of order $Y_{1,2}\sim 10^{-2}$. So assuming Yukawa couplings of this order, for $\mu\sim 1$ TeV, one can get $m_{\nu}\sim 10^{-10}$ GeV for $r_{ca}\sim 40$. However, this choice is not unique and presented only for a demonstration for natural values of the Yukawa couplings.

To the best of the author's knowledge, the presented model in this work is the only \textit{true two-loop} neutrino mass in the context of $\bm{SU(5)}$ GUT.  Note that, one can think of embedding the original two-loop Zee-Babu mechanism \cite{Babu:1988ki} (right diagram in Fig. \ref{zee-babu}) in $\bm{SU(5)}$ GUT. 
Zee-Babu mechanism requires a singly charged singlet $(1,1,1)$ and a doubly charged singlet $(1,1,2)$ under the SM. These multiplets can be embedded in $(1,1,1) \subset \bm{10_H}$  and $(1,1,-2) \subset \bm{50_H}$ representations of $\bm{SU(5)}$. Introduction of  $\bm{10_H}$  brings new Yukawa couplings into the theory contained in the following   bilinear:
\begin{align}
\bm{\overline{5}}\times \bm{\overline{5}} = \bm{\overline{10}} + \bm{\overline{15}}.
\end{align}

\noindent Hence the requirement of both $\mathcal{L}\supset Y_5 \bm{10_F} \bm{10_F} \bm{50_H}$ and $\mathcal{L}\supset Y_6 \bm{\overline{5}_F} \bm{\overline{5}_F} \bm{10_H}$ Yukawa couplings into the theory are required, where  $Y_5$ is a symmetric $3\times 3$ matrix where as $Y_6$ is  anti-symmetric $3\times 3$ matrix in the flavor space. 
However, presence of the Yukawa coupling $Y_6$ and the allowed gauge invariant cubic term $\mu \; \bm{5_H}\bm{5_H}\bm{10^{\ast}_H}$ in the scalar potential automatically leads to one-loop diagram  via Zee-mechanism \cite{Wolfenstein:1980sf}   (left diagram in Fig. \ref{zee-babu}). 
This is why, such an embedding which was realized in \cite{Tamvakis:1985if},  cannot be a \textit{true two-loop} neutrino mass model.  Similar conclusion can be reached for the $\bm{SU(5)}$ model presented in \cite{Wu:1980hz}, due to the presence of $\bm{10_H}$ Higgs, in addition to their two-loop diagram, one-loop diagram of the Zee-type automatically appears.  So the model presented in this work is unique in its features.

%%%%%%%%%%%%%%%%%%%%%%%%%%%%%%%%%%%%%%%%%%%%%%%
%%%%%%%%%%%%%%%%%%%%%%%%%%%%%%%%%%%%%%%%%%%%%%%
\section{Scalar potential and the Higgs Bosons mass spectrum}\label{SEC-04}
As aforementioned, the minimal model consists of the Higgs set $\bm{24_H}$, $\bm{5_H}$,  and $\bm{45_H}$. However, this model is still defective since neutrinos remain massless. In the previous section it is shown that  to build a \textit{true two-loop} neutrino mass model the minimal model needs to be extended by two more Higgs multiplet $\bm{40_H}$ and $\bm{50_H}$.
 In this section, we compute the Higgs mass spectrum of the set $\bm{5_H}+\bm{24_H}+\bm{40_H}+\bm{45_H}+\bm{50_H}$ after the GUT symmetry is broken spontaneously.  This analysis is performed to find the Higgs mass relationships,  which  will be relevant for our study of the gauge coupling unification performed in the next section. The tensorial properties of all the Higgs multiplets in our framework   are presented in table \ref{fields}.

\FloatBarrier
\begin{table}[th]  
\centering
\begin{tabular}{|c|c|c|}
\hline
fields &notation&properties   \\ \hline \hline
$\bm{24_H}$ &$\bm{\Phi}_{\alpha}^{\beta}$& adjoint 24-dimensional real traceless, $\bm{\Phi}_{\alpha}^{\alpha} =0$  \\ \hline\hline
$\bm{5_H}$ &$\bm{\phi}_{\alpha}$& fundamental 5-dimensional complex\\ \hline \hline 
$\bm{45_H}$ &$\bm{\Sigma}_{\alpha \beta}^{\gamma}$&   
\pbox{20cm}{\vspace{2pt}
45-dimensional complex, anti-symmetric under, $\bm{\Sigma}_{\alpha \beta}^{\gamma}=-\bm{\Sigma}_{\beta \alpha }^{\gamma}$  \\  and traceless,  $\bm{\Sigma}_{\alpha \beta}^{\alpha}=0$
\vspace{2pt}} \\ \hline\hline
$\bm{40_H}$ &$\bm{\eta}^{\alpha \beta \gamma}$& 
\pbox{20cm}{\vspace{2pt}
40-dimensional complex, anti-symmetric under, $\bm{\eta}^{\alpha \beta \gamma}=-\bm{\eta}^{\beta \alpha \gamma}$ \\ and and $\bm{\eta}^{\alpha \beta \gamma}\epsilon_{\alpha \beta \gamma \rho \sigma}=0$
\vspace{2pt}} \\ \hline\hline
$\bm{50_H}$ &$\bm{\chi}^{\alpha \beta \gamma \delta}$& 
\pbox{20cm}{\vspace{2pt}
50-dimensional complex, symmetric under, $\bm{\chi}^{\alpha \beta \gamma \delta}=\bm{\chi}^{\gamma \delta \alpha \beta }$ \\
anti symmetric under,  $\bm{\chi}^{\alpha \beta \gamma \delta}=-\bm{\chi}^{ \beta \alpha \gamma \delta  }=- \bm{\chi}^{\alpha \beta  \delta \gamma}$ \\
and additionally $\bm{\chi}^{\alpha \beta \gamma \delta} \;\epsilon_{\xi \alpha \beta \gamma \delta}=0$ 
\vspace{2pt}}\\ [0.5ex] \hline
\end{tabular} 
\caption{ Particle content of our model and their relevant properties.} \label{fields}
\end{table}

We are interested in the mass spectrum of the Higgs bosons as a result of breaking of the GUT symmetry  down to the SM gauge group. The only field that acquires VEV at this stage is the $\bm{24_H}$ as a result,  the mass of the Higgs multiplets come from the interaction with the $\bm{24_H}$ representation. For our purpose of this  analysis, the effect of the EW scale VEVs of the $\bm{5_H}$ and $\bm{45_H}$ fields can be completely ignored. Then the relevant part of the scalar potential contributing to their masses is given by: 
\begin{align}
V= V_{24}+V_{24,5}+V_{24,45}+V_{24,40}+V_{24,50}+V_{mix},
\end{align} 

\noindent where,
\begin{align}
V_{24}
= 
\frac{1}{2} m^2_{24}\; \Phi_{\alpha}^{\beta} \Phi_{\beta}^{\alpha} 
+
\mu_{24}\; \Phi_{\alpha}^{\beta} \Phi_{\beta}^{\gamma} \Phi_{\gamma}^{\alpha} 
+ 
\lambda_1\; (\Phi_{\alpha}^{\beta} \Phi_{\beta}^{\alpha})^2 
+ 
\lambda_2\; \Phi_{\alpha}^{\beta} \Phi_{\beta}^{\gamma} \Phi_{\gamma}^{\delta}  \Phi_{\delta}^{\alpha}\;,
\end{align}
\vspace{-20pt}
\begin{align}
V_{24,5}
= 
m^2_{5}\; \phi_{\alpha} \phi^{\ast \alpha} 
+
\mu_{5}\; \phi_{\alpha} \phi^{\ast \beta} \Phi_{\beta}^{\alpha} 
+
 \alpha_1\; (\phi_{\alpha} \phi^{\ast \alpha}) ( \Phi_{\gamma}^{\beta} \Phi_{\beta}^{\gamma}) 
 + 
 \alpha_2\; \phi_{\alpha} \phi^{\ast \beta} \Phi_{\beta}^{\gamma} \Phi_{\gamma}^{\alpha}\;,
\end{align}
\vspace{-20pt}
\begin{align}
V_{24,45}
&= 
m^2_{45}\; \Sigma_{\alpha\beta}^{\gamma} \Sigma^{\ast \alpha\beta}_{\gamma} +\mu_{45}\;  \Sigma_{\alpha\beta}^{\gamma} \Sigma^{\ast \alpha\beta}_{\delta} \Phi_{\gamma}^{\delta}  
+
\mu_{45}^{\prime}\;  \Sigma_{\alpha\beta}^{\gamma} \Sigma^{\ast \beta\delta}_{\gamma} \Phi_{\delta}^{\alpha}
\nonumber \\ &
+ 
\xi_1\; (\Sigma_{\alpha\beta}^{\gamma} \Sigma^{\ast \alpha\beta}_{\gamma}) (\Phi_{\rho}^{\sigma} \Phi_{\sigma}^{\rho})
+ 
\xi_2\; \Sigma_{\alpha\beta}^{\gamma} \Sigma^{\ast \alpha\beta}_{\delta} \Phi_{\gamma}^{\rho} \Phi_{\rho}^{\delta}
+ 
\xi_3\; \Sigma_{\alpha\beta}^{\gamma} \Sigma^{\ast \alpha\delta}_{\gamma} \Phi_{\delta}^{\rho} \Phi_{\rho}^{\beta}
\nonumber \\ &
+ 
\xi_4\; \Sigma_{\alpha\beta}^{\gamma} \Sigma^{\ast \alpha\delta}_{\rho} \Phi_{\gamma}^{\beta} \Phi_{\delta}^{\rho}
+ 
\xi_5\; \Sigma_{\alpha\beta}^{\gamma} \Sigma^{\ast \alpha\delta}_{\rho} \Phi_{\gamma}^{\rho} \Phi_{\delta}^{\beta}
+ 
\xi_6\; \Sigma_{\alpha\beta}^{\gamma} \Sigma^{\ast \delta\rho}_{\gamma} \Phi_{\delta}^{\alpha} \Phi_{\rho}^{\beta}\;,
\end{align}
\vspace{-20pt}
\begin{align}
V_{24,40}
&= 
m^2_{40}\; \eta^{\alpha\beta\gamma} \eta^{\ast}_{\alpha\beta\gamma}
+
\mu_{40}\; \eta^{\alpha\beta\gamma} \eta^{\ast}_{\alpha\beta\delta} \Phi_{\gamma}^{\delta}
+
\mu_{40}^{\prime}\; \eta^{\alpha\beta\gamma} \eta^{\ast}_{\delta\beta\gamma} \Phi_{\alpha}^{\delta}
+
\omega_1 (\eta^{\alpha\beta\gamma} \eta^{\ast}_{\alpha\beta\gamma}) (\Phi_{\rho}^{\sigma} \Phi_{\sigma}^{\rho})
+
\omega_2 \eta^{\alpha\beta\gamma} \eta^{\ast}_{\alpha\beta\delta} \Phi_{\gamma}^{\sigma} \Phi_{\sigma}^{\delta}
\nonumber \\ &
+
\omega_3 \eta^{\alpha\beta\gamma} \eta^{\ast}_{\alpha\delta\gamma} \Phi_{\beta}^{\sigma} \Phi_{\sigma}^{\delta}
+
\omega_4 \eta^{\alpha\beta\gamma} \eta^{\ast}_{\delta\rho\gamma} \Phi_{\alpha}^{\delta} \Phi_{\beta}^{\rho}
+
\omega_5 \eta^{\alpha\beta\gamma} \eta^{\ast}_{\delta\beta\rho} \Phi_{\gamma}^{\delta} \Phi_{\gamma}^{\rho}\;,
\end{align}
\vspace{-20pt}
\begin{align}
V_{24,50}
&= 
m^2_{50}\; \chi^{\alpha\beta\gamma\delta} \chi^{\ast}_{\alpha\beta\gamma\delta}
+
\mu_{50}\; \chi^{\alpha\beta\gamma\delta} \chi^{\ast}_{\rho\beta\gamma\alpha} \Phi_{\delta}^{\rho}
\nonumber \\ &
+
\zeta_1 (\chi^{\alpha\beta\gamma\delta} \chi^{\ast}_{\alpha\beta\gamma\delta}) (\Phi_{\rho}^{\sigma} \Phi_{\sigma}^{\rho})
+
\zeta_2 \chi^{\alpha\beta\gamma\delta} \chi^{\ast}_{\delta\gamma\alpha\beta} \Phi_{\delta}^{\sigma} \Phi_{\sigma}^{\rho}
+
\zeta_3 \chi^{\alpha\beta\gamma\delta} \chi^{\ast}_{\gamma\delta\rho\sigma} \Phi_{\alpha}^{\rho} \Phi_{\beta}^{\sigma}\;,
\end{align}
\vspace{-20pt}
\begin{align}\label{mix}   
&V_{mix}=
 \mu_1\; \Sigma_{\alpha\beta}^{\gamma} \Phi_{\gamma}^{\alpha} \phi^{\ast \beta}
 +
\kappa_1\;  \Sigma_{\alpha\beta}^{\gamma} \Phi_{\gamma}^{\alpha} \Phi_{\delta}^{\beta} \phi^{\ast \delta}  
 +
\kappa_2\;  \Sigma_{\alpha\beta}^{\gamma} \Phi_{\gamma}^{\delta} \Phi_{\delta}^{\alpha} \phi^{\ast \beta}  
\nonumber \\ &
+ 
\mu_2\;\Phi_{\alpha}^{\beta} \chi^{\rho\sigma\alpha\delta} \Sigma_{\delta}^{\ast \tau\kappa} \epsilon_{\rho\sigma\tau\kappa\beta} 
 +
 \kappa_3\;\Phi_{\rho}^{\alpha}\Phi_{\sigma}^{\beta}\phi_{\nu}
 \chi^{\ast}_{\alpha\beta\tau\kappa}\epsilon^{\rho\sigma\tau\kappa\nu}
 +
  \kappa_4\;\Phi_{\alpha}^{\beta}\Phi_{\beta}^{\gamma}
 \chi^{\rho\sigma\alpha\delta}\Sigma^{\ast\tau\kappa}_{\delta}
 \epsilon_{\rho\sigma\tau\kappa\gamma}
+h.c. \;.
\end{align}

\noindent 
The SM singlet component, $\Phi_1(1,1,0)$ of the adjoint Higgs acquires VEV, $\langle \bm{\Phi_H} \rangle \equiv V_{GUT}$ and breaks the GUT symmetry down to the SM group. The minimization condition demands:
\begin{align}
m^2_{24}= -\lambda _1 V_{\text{GUT}}^2-\frac{7}{30} \lambda _2 V_{\text{GUT}}^2-\frac{1}{4}
   \sqrt{\frac{3}{5}} \mu _{24} V_{\text{GUT}}.
\end{align}

\noindent
The multiplets $(3,2,-\frac{5}{6})$ and $(3,2,\frac{5}{6})$ from  $\bm{24_H}$ field correspond to the Goldstone bosons and hence eaten up by the massive gauge bosons.  The masses of the other multiplets in $\bm{24_H}$ are given by:
\begin{align}
&m^2_{\Phi_1}= \frac{1}{60} V_{\text{GUT}} \left(120 \lambda _1 V_{\text{GUT}}+28 \lambda _2 V_{\text{GUT}}+3 \sqrt{15} \mu
   _{24}\right),
\\
&m^2_{\Phi_2}= \frac{1}{12} V_{\text{GUT}} \left(8 \lambda _2 V_{\text{GUT}}+3 \sqrt{15} \mu _{24}\right),
   \\
&m^2_{\Phi_3}= \frac{1}{12} V_{\text{GUT}} \left(2 \lambda _2 V_{\text{GUT}}-3 \sqrt{15} \mu _{24}\right).
\end{align} 

\noindent
The mass spectrum of the multiplets residing in $\bm{5_H}$ Higgs by neglecting  mixing with other fields are given by:
\begin{align}
&m^2_{\phi_1}=
\frac{1}{30} \left(15 \alpha _1 V_{\text{GUT}}^2+2 \alpha _2 V_{\text{GUT}}^2-2 \sqrt{15} \mu _5 V_{\text{GUT}}+30
   m^2_5\right),
   \\
&m^2_{\phi_2}=   \frac{1}{30} \left(15 \alpha _1 V_{\text{GUT}}^2+2 \alpha _2 V_{\text{GUT}}^2-2 \sqrt{15} \mu _5 V_{\text{GUT}}+30
   m^2_5\right).
\end{align}  

\noindent Similarly, ignoring the mixing of the fields, the 
mass spectrum of the multiplets contained in the $\bm{45_H}$ Higgs are given by:
\begin{align}
m^2_{\Sigma_1}&=
\frac{1}{480} \left(28 \sqrt{15} \mu _{45} V_{\text{GUT}}-38 \sqrt{15} \mu_{45}^{\prime} V_{\text{GUT}}+240 \xi _1
   V_{\text{GUT}}^2+62 \xi _2 V_{\text{GUT}}^2+67 \xi _3 V_{\text{GUT}}^2+75 \xi _4 V_{\text{GUT}}^2
   \nonumber  \right. \\& \left.   
   +52 \xi _5
   V_{\text{GUT}}^2
+42 \xi _6 V_{\text{GUT}}^2+480 m^2_{45}\right),
\\
m^2_{\Sigma_2}&=
\frac{1}{240} \left(4 \sqrt{15} \mu _{45} V_{\text{GUT}}+6 \sqrt{15} \mu_{45}^{\prime} V_{\text{GUT}}+120 \xi _1
   V_{\text{GUT}}^2+26 \xi _2 V_{\text{GUT}}^2+21 \xi _3 V_{\text{GUT}}^2+50 \xi _4 V_{\text{GUT}}^2
   \nonumber  \right. \\& \left.   
   +11 \xi _5
   V_{\text{GUT}}^2
-4 \xi _6 V_{\text{GUT}}^2+240 m^2_{45}\right),
\\
m^2_{\Sigma_3}&=
\frac{1}{60} \left(-4 \sqrt{15} \mu _{45} V_{\text{GUT}}-6 \sqrt{15} \mu_{45}^{\prime} V_{\text{GUT}}+30 \xi _1
   V_{\text{GUT}}^2+4 \xi _2 V_{\text{GUT}}^2+9 \xi _3 V_{\text{GUT}}^2-6 \xi _5 V_{\text{GUT}}^2
   \nonumber  \right. \\& \left.   
+9 \xi _6
   V_{\text{GUT}}^2+60 m^2_{45}\right),
\\
m^2_{\Sigma_4}&=
\frac{1}{60} \left(6 \sqrt{15} \mu _{45} V_{\text{GUT}}+4 \sqrt{15} \mu_{45}^{\prime} V_{\text{GUT}}+30 \xi _1
   V_{\text{GUT}}^2+9 \xi _2 V_{\text{GUT}}^2+4 \xi _3 V_{\text{GUT}}^2-6 \xi _5 V_{\text{GUT}}^2
   \nonumber  \right. \\& \left.   
+4 \xi _6
   V_{\text{GUT}}^2+60 m^2_{45}\right),
\\
m^2_{\Sigma_5}&=
\frac{1}{120} \left(12 \sqrt{15} \mu _{45} V_{\text{GUT}}-2 \sqrt{15} \mu_{45}^{\prime} V_{\text{GUT}}+60 \xi _1
   V_{\text{GUT}}^2+18 \xi _2 V_{\text{GUT}}^2+13 \xi _3 V_{\text{GUT}}^2+3 \xi _5 V_{\text{GUT}}^2
   \nonumber  \right. \\& \left.   
-12 \xi _6
   V_{\text{GUT}}^2+120 m^2_{45}\right),
\\
m^2_{\Sigma_6}&=
\frac{1}{30} \left(-2 \sqrt{15} \mu _{45} V_{\text{GUT}}+2 \sqrt{15} \mu_{45}^{\prime} V_{\text{GUT}}+15 \xi _1
   V_{\text{GUT}}^2+2 \xi _2 V_{\text{GUT}}^2+2 \xi _3 V_{\text{GUT}}^2+2 \xi _5 V_{\text{GUT}}^2
   \nonumber  \right. \\& \left.   
+2 \xi _6
   V_{\text{GUT}}^2+30 m^2_{45}\right),
\\
m^2_{\Sigma_7}&=
\frac{1}{120} \left(-8 \sqrt{15} \mu _{45} V_{\text{GUT}}-2 \sqrt{15} \mu_{45}^{\prime} V_{\text{GUT}}+60 \xi _1
   V_{\text{GUT}}^2+8 \xi _2 V_{\text{GUT}}^2+13 \xi _3 V_{\text{GUT}}^2-2 \xi _5 V_{\text{GUT}}^2
   \nonumber  \right. \\& \left.   
-12 \xi _6
   V_{\text{GUT}}^2+120 m^2_{45}\right).
\end{align}

\noindent 
Mass spectrum of the multiplets of $\bm{40_H}$ field:
\begin{align}
m^2_{\eta_1}&= 
\frac{1}{20} \left(2 \sqrt{15} \mu _{40} V_{\text{GUT}}+2 \sqrt{15} \mu_{40}^{\prime} V_{\text{GUT}}+10 \omega _1
   V_{\text{GUT}}^2+3 \omega _2 V_{\text{GUT}}^2+3 \omega _3 V_{\text{GUT}}^2
\nonumber  \right. \\& \left.       
+3 \omega _4 V_{\text{GUT}}^2+3 \omega
   _5 V_{\text{GUT}}^2+20 m^2_{40}\right),
\\
m^2_{\eta_2}&= 
\frac{1}{360} \left(-4 \sqrt{15} \mu _{40} V_{\text{GUT}}+26 \sqrt{15} \mu_{40}^{\prime} V_{\text{GUT}}+180 \omega
   _1 V_{\text{GUT}}^2+34 \omega _2 V_{\text{GUT}}^2+49 \omega _3 V_{\text{GUT}}^2
\nonumber  \right. \\& \left.       
+24 \omega _4 V_{\text{GUT}}^2-21
   \omega _5 V_{\text{GUT}}^2+360 m^2_{40}\right),
\\
m^2_{\eta_3}&= 
\frac{1}{360} \left(16 \sqrt{15} \mu _{40} V_{\text{GUT}}-14 \sqrt{15} \mu_{40}^{\prime} V_{\text{GUT}}+180 \omega
   _1 V_{\text{GUT}}^2+44 \omega _2 V_{\text{GUT}}^2+29 \omega _3 V_{\text{GUT}}^2
\nonumber  \right. \\& \left.       
+4 \omega _4 V_{\text{GUT}}^2-26
   \omega _5 V_{\text{GUT}}^2+360 m^2_{40}\right),
\\
m^2_{\eta_4}&= 
\frac{1}{120} \left(12 \sqrt{15} \mu _{40} V_{\text{GUT}}+2 \sqrt{15} \mu_{40}^{\prime} V_{\text{GUT}}+60 \omega _1
   V_{\text{GUT}}^2+18 \omega _2 V_{\text{GUT}}^2+13 \omega _3 V_{\text{GUT}}^2
\nonumber  \right. \\& \left.       
-12 \omega _4 V_{\text{GUT}}^2+3
   \omega _5 V_{\text{GUT}}^2+120 m^2_{40}\right),
\\
m^2_{\eta_5}&= 
\frac{1}{120} \left(-8 \sqrt{15} \mu _{40} V_{\text{GUT}}+2 \sqrt{15} \mu_{40}^{\prime} V_{\text{GUT}}+60 \omega _1
   V_{\text{GUT}}^2+8 \omega _2 V_{\text{GUT}}^2+13 \omega _3 V_{\text{GUT}}^2
\nonumber  \right. \\& \left.       
-12 \omega _4 V_{\text{GUT}}^2-2 \omega
   _5 V_{\text{GUT}}^2+120 m^2_{40}\right),
\\
m^2_{\eta_6}&= 
\frac{1}{30} \left(-2 \sqrt{15} \mu _{40} V_{\text{GUT}}-2 \sqrt{15} \mu_{40}^{\prime} V_{\text{GUT}}+15 \omega _1
   V_{\text{GUT}}^2+2 \omega _2 V_{\text{GUT}}^2+2 \omega _3 V_{\text{GUT}}^2
\nonumber  \right. \\& \left.       
+2 \omega _4 V_{\text{GUT}}^2+2 \omega
   _5 V_{\text{GUT}}^2+30 m^2_{40}\right).
\end{align}

\noindent 
And finally the mass spectrum of the multiplets residing in $\bm{50_H}$ field:
\begin{align}
m^2_{\chi_1}&= \frac{1}{20} \left(10 \zeta  V_{\text{GUT}}^2-3 \zeta_1 V_{\text{GUT}}^2+3 \zeta_2
   V_{\text{GUT}}^2+\sqrt{15} \mu _{50} V_{\text{GUT}}+20 m^2_{50}\right),
\\
m^2_{\chi_2}&= \frac{1}{360} \left(180 \zeta  V_{\text{GUT}}^2-39 \zeta_1 V_{\text{GUT}}^2+14 \zeta_2
   V_{\text{GUT}}^2+3 \sqrt{15} \mu _{50} V_{\text{GUT}}+360 m^2_{50}\right),
\\
m^2_{\chi_3}&= \frac{1}{240} \left(120 \zeta  V_{\text{GUT}}^2-31 \zeta_1 V_{\text{GUT}}^2+6 \zeta_2 V_{\text{GUT}}^2+7
   \sqrt{15} \mu _{50} V_{\text{GUT}}+240 m^2_{50}\right),
\\
m^2_{\chi_4}&= \frac{1}{30} \left(15 \zeta  V_{\text{GUT}}^2-2 \zeta_1 V_{\text{GUT}}^2+2 \zeta_2
   V_{\text{GUT}}^2-\sqrt{15} \mu _{50} V_{\text{GUT}}+30 m^2_{50}\right),
\\
m^2_{\chi_5}&= \frac{1}{120} \left(60 \zeta  V_{\text{GUT}}^2-13 \zeta_1 V_{\text{GUT}}^2-12 \zeta_2
   V_{\text{GUT}}^2+\sqrt{15} \mu _{50} V_{\text{GUT}}+120 m^2_{50}\right),
\\
m^2_{\chi_6}&= 
\frac{1}{240} \left(120 \zeta  V_{\text{GUT}}^2-21 \zeta_1 V_{\text{GUT}}^2-4 \zeta_2
   V_{\text{GUT}}^2-3 \sqrt{15} \mu _{50} V_{\text{GUT}}+240 m^2_{50}\right).
\end{align}

These mass spectrum helps one to understand whether splitting among different multiplets originating from the same field is possible or not. Splitting among different multiplets  for some of the fields  is necessary to achieve unification  to be discussed in the next section.  From the mass spectrum computed above, it can be realized that, due to enough number of parameters, there is no mass relationship among the multiplets of $\bm{40_H}$. This is also true for $\bm{5_H}$, $\bm{24_H}$ and $\bm{45_H}$.  However, which is not true for  the multiplets contained in $\bm{50_H}$ and from the above calculation we find:
\begin{align}
&m^2_{\chi_4}=3m^2_{\chi_2}-2m^2_{\chi_3},
\\
&m^2_{\chi_5}=2m^2_{\chi_3}-m^2_{\chi_1},
\\
&m^2_{\chi_6}=\frac{3}{2}m^2_{\chi_2}-\frac{1}{2}m^2_{\chi_1}.
\end{align}

\noindent 
For the study of the gauge coupling unification, we impose the mass relations as derived above.

Till now, we have ignored the mixings among  the multiplets having the same quantum number  coming from different Higgs  representations. For completeness  here we take into account such mixings.  Note that the relevant mixing terms are contained in the $V_{mix}$ term given in Eq. \eqref{mix}. Now taking these mixed  terms into consideration,    the mixing between the iso-spin doublets, $(1,2,\frac{1}{2})$  present in $\bm{5_H}$ and $\bm{45_H}$ representations are given by:
\begin{align}\label{doublet}
\begin{pmatrix}
\phi^{(D)}_1&\Sigma^{(D)}_1
\end{pmatrix}
\begin{pmatrix}
m^2_{\phi_1}&m^2_{D_{12}}\\
m^2_{D_{12}}&m^2_{\Sigma_1}
\end{pmatrix}
\begin{pmatrix}
\phi^{(D)\ast}_1 \\ \Sigma^{(D)\ast}_1
\end{pmatrix},
\end{align}

\noindent with,
\begin{align}
&m^2_{D_{12}}=\frac{1}{24\sqrt{2}} V_{\text{GUT}} \left(3 \sqrt{3} \kappa _1 V_{\text{GUT}}+\sqrt{3} \kappa _2 V_{\text{GUT}}+6 \sqrt{5} \mu_1
   \right)  . 
\end{align}

\noindent
Furthermore, the color triplets, $(3,1,-1/3)$ contained in $\bm{5_H}$, 
$\bm{45_H}$ and $\bm{50_H}$ mix with each and we find: 
\begin{align}\label{triplet}
\begin{pmatrix}
\phi^{(T)}_2&\Sigma^{(T)}_2&\chi^{(T)}_2
\end{pmatrix}
\begin{pmatrix}
m^2_{\phi_2}&m^2_{T_{12}}&m^2_{T_{13}}\\
m^2_{T_{12}}&m^2_{\Sigma_2}&m^2_{T_{23}}\\
m^2_{T_{13}}&m^2_{T_{23}}&m^2_{\chi_2}
\end{pmatrix}
\begin{pmatrix}
\phi^{(T)\ast}_2 \\ \Sigma^{(T)\ast}_2 \\ \chi^{(T)\ast}_2
\end{pmatrix},
\end{align}

\noindent with,
\begin{align}
&m^2_{T_{12}}=-\frac{1}{12\sqrt{2}} V_{\text{GUT}} \left(2  \kappa _1 V_{\text{GUT}}-\kappa _2 V_{\text{GUT}}-2 \sqrt{15} \mu_1
   \right)   ,
   \\ &
   m^2_{T_{13}}=
   -\frac{5}{6\sqrt{3}}\kappa_3V_{\text{GUT}}^2,
   \\ &
   m^2_{T_{23}}= 
   -\frac{1}{9\sqrt{2}}V_{\text{GUT}} (6\sqrt{5}\mu_2+\sqrt{3}\kappa_4V_{\text{GUT}}).
\end{align}

Note that, due to the breaking of the GUT symmetry, all the multiplets acquire mass   of the order of the GUT scale. However, to break the SM symmetry, the SM like Higgs doublet needs to be kept at the EW scale. This can be achieved by imposing the well known  fine-tuning condition in the doublet mass matrix Eq. \eqref{doublet}. This fine-tuning does not leave any color triplet Higgs light that can be seen from the corresponding mass matrix given in Eq. \eqref{triplet}.

%%%%%%%%%%%%%%%%%%%%%%%%%%%%%%%%%%%%%%%%%%%%%%%
%%%%%%%%%%%%%%%%%%%%%%%%%%%%%%%%%%%%%%%%%%%%%%%
\section{Gauge coupling unification and proton decay constraints}\label{SEC-05}
In this section we present few different scenarios where successful gauge coupling unification within our framework   can be achieved  which are also  in agreement with the proton decay bounds.  For the gauge couplings the renormalization group equations can be written as:
\begin{align}\label{RGE}
 \alpha_i^{-1}(M_Z) = \alpha^{-1}_{GUT}+\frac{B_i}{2\pi} \text{ln}\left(\frac{M_{GUT}}{M_Z}\right),
\end{align}

where,
\begin{align}
&b_i^{SM}=(\frac{41}{10}, -\frac{19}{6}, -7),
\\
&B_i=b_i^{SM} + \Delta b_{i,k} r_k,
\\
&r_k= \frac{\text{ln} {(M_{GUT} / M_k)}}{\text{ln} {(M_{GUT} / M_Z)}}.
\end{align}

\noindent
Here, $b_i^{SM}$ are the SM $\beta$-coefficients and $r_k$ represents the threshold weight factor of the BSM multiplet $k$ of mass   
 $M_k$. To affect the coupling running, the BSM multiplet $k$ needs to    live in a scale that is in between the  electroweak scale and the GUT scale, here we assume that the rest of the mutiplets are degenerate with the unification scale. $\Delta b_{i,k}=b_{i,k}-b_{i,k-1}$ is the increase in the RGE coefficient at the threshold, $M_k$ for a BSM multiplet. It is convenient to rewrite the equations for the running of the gauge couplings in terms of the low energy observables at the electroweak scale  and the differences in the coefficients 
$B_{ij}=B_i-B_j$ \cite{Giveon:1991zm}. In this way, the equations become:
\begin{align}
& \frac{B_{23}}{B_{12}} =\frac{5}{8}\left(\frac{\text{sin}^2\theta_W (M_Z) - \alpha (M_Z) / \alpha_s (M_Z)}{ 3/8-\text{sin}^2\theta_W (M_Z)}\right), \nonumber \\
 \label{ratio}\\
& \displaystyle \text{ln}\left(\frac{M_{GUT}}{M_Z}\right)=\frac{16\pi}{ 5\alpha (M_Z)}\left(\frac{3/8-\text{sin}^2\theta_W (M_Z)}{ B_{12}}\right). 
\end{align}
\\
From the  the experimental measurements, $\alpha (M_Z)^{-1}=127.94$, $\sin^2 \theta_W (M_Z)=0.231$, and $\alpha_s (M_Z)=0.1185$ \cite{Agashe:2014kda} which infers 
\begin{align}
& \frac{B_{23}}{B_{12}}= 0.718, \\
& M_{GUT}= M_Z\; exp\left(\frac{184.87}{B_{12}}\right).\label{MGUT}
\end{align}

\FloatBarrier
\begin{table}[b!]

\begin{minipage}{.5\linewidth}
\centering
\begin{tabular}{|c|c|c|}
\hline
fields &$\Delta B_{12}$&$\Delta B_{23}$   \\ \hline\hline

%%%%%%%%%%%%%%%%%%%%%%%%%%%%%%%%%%%%%%%%%%%%
$\Phi_2 (1,3,0)$&$-\frac{1}{3}r_{\Phi_2}$&$\frac{1}{3}r_{\Phi_2}$\\
$\Phi_3 (8,1,0)$&$0$&$-\frac{1}{2}r_{\Phi_3}$\\ \hline\hline

%%%%%%%%%%%%%%%%%%%%%%%%%%%%%%%%%%%%%%%%%%%%
$\phi_1 (1,2,\frac{1}{2})$&$-\frac{1}{15}r_{\phi_1}$&$\frac{1}{6}r_{\phi_1}$\\
$\phi_2 (3,1,-\frac{1}{3})$&$\frac{1}{15}r_{\phi_2}$&$-\frac{1}{6}r_{\phi_2}$\\ \hline\hline

%%%%%%%%%%%%%%%%%%%%%%%%%%%%%%%%%%%%%%%%%%%%
$\Sigma_1 (1,2,\frac{1}{2})$&$-\frac{1}{15}r_{\Sigma_1}$&$\frac{1}{6}r_{\Sigma_1}$\\
$\Sigma_2 (3,1,-\frac{1}{3})$&$\frac{1}{15}r_{\Sigma_2}$&$-\frac{1}{6}r_{\Sigma_2}$\\ 
$\Sigma_3 (\overline{3},1,\frac{4}{3})$&$\frac{16}{15}r_{\Sigma_3}$&$-\frac{1}{6}r_{\Sigma_3}$\\
$\Sigma_4 (\overline{3},2,-\frac{7}{6})$&$\frac{17}{15}r_{\Sigma_4}$&$\frac{1}{6}r_{\Sigma_4}$\\
$\Sigma_5 (3,3,-\frac{1}{3})$&$-\frac{9}{5}r_{\Sigma_5}$&$\frac{3}{2}r_{\Sigma_5}$\\
$\Sigma_6 (\overline{6},1,-\frac{1}{3})$&$\frac{2}{15}r_{\Sigma_6}$&$-\frac{5}{6}r_{\Sigma_6}$\\
$\Sigma_7 (8,2,\frac{1}{2})$&$-\frac{8}{15}r_{\Sigma_7}$&$-\frac{2}{3}r_{\Sigma_7}$\\ \hline
\end{tabular} 

\end{minipage}%
\begin{minipage}{.5\linewidth}
    
\centering
\begin{tabular}{|c|c|c|}
\hline
fields &$\Delta B_{12}$&$\Delta B_{23}$   \\ \hline\hline

%%%%%%%%%%%%%%%%%%%%%%%%%%%%%%%%%%%%%%%%%%%%
$\eta_1 (1,2,-\frac{3}{2})$&$\frac{11}{15}r_{\eta_1}$&$\frac{1}{6}r_{\eta_1}$\\
$\eta_2 (\overline{3},1,-\frac{2}{3})$&$\frac{4}{15}r_{\eta_2}$&$-\frac{1}{6}r_{\eta_2}$\\ 
$\eta_3 (3,2,\frac{1}{6})$&$-\frac{7}{15}r_{\eta_3}$&$\frac{1}{6}r_{\eta_3}$\\
$\eta_4 (\overline{3},3,-\frac{2}{3})$&$-\frac{6}{5}r_{\eta_4}$&$\frac{3}{2}r_{\eta_4}$\\
$\eta_5 (\overline{6},2,\frac{1}{6})$&$-\frac{14}{15}r_{\eta_5}$&$-\frac{2}{3}r_{\eta_5}$\\
$\eta_6 (8,1,1)$&$\frac{8}{5}r_{\eta_6}$&$-r_{\eta_6}$\\ \hline\hline

%%%%%%%%%%%%%%%%%%%%%%%%%%%%%%%%%%%%%%%%%%%%  
$\chi_1 (1,1,-2)$&$\frac{4}{5}r_{\chi_1}$&$0$\\
$\chi_2 (3,1,-\frac{1}{3})$&$\frac{1}{15}r_{\chi_2}$&$-\frac{1}{6}r_{\chi_2}$\\ 
$\chi_3 (\overline{3},2,-\frac{7}{6})$&$\frac{17}{15}r_{\chi_3}$&$\frac{1}{6}r_{\chi_3}$\\
$\chi_4 (6,1,\frac{4}{3})$&$\frac{32}{15}r_{\chi_4}$&$-\frac{5}{6}r_{\chi_4}$\\
$\chi_5 (\overline{6},3,-\frac{1}{3})$&$-\frac{18}{5}r_{\chi_5}$&$\frac{3}{2}r_{\chi_5}$\\
$\chi_6 (8,2,\frac{1}{2})$&$-\frac{8}{15}r_{\chi_6}$&$-\frac{2}{3}r_{\chi_6}$
\\ \hline 
\end{tabular} 

\end{minipage} 
\caption{ $B_{ij}$ coefficients of the multiplets present in our theory.}\label{Bij}
\end{table}

So to achieve unification, $\frac{B_{23}}{B_{12}}$ ratio needs to be $0.718\pm 0.005$ ($\pm 1\sigma$ range), whereas in the SM model this ratio is equal to 0.528, hence fails badly to unify gauge couplings. The corresponding scale for the SM from Eq. \eqref{MGUT} is found to be $10^{13}$ GeV.  So threshold corrections  from the BSM multiplets is needed to modify the $B_{12}$ and $B_{23}$ to get the correct ratio. These $B_{ij}$ coefficients for all the BSM multiplets for our model with $\bm{5_H}+ \bm{24_H}+ \bm{40_H}+ \bm{45_H}+ \bm{50_H}$ are presented in table. \ref{Bij} and are used in our study of the gauge coupling unification.

The main experimental test of the existence of GUTs is via the detection of the  proton decay yet to be observed. In GUT models in the non-supersymmetric framework, the leading contribution to the proton decay is due to  the gauge mediated $d=6$  operators. In $\bm{SU(5)}$ GUT, the gauge bosons that are responsible for the proton to decay are $(3,2,-\frac{5}{6}) + (\overline{3},2,\frac{5}{6})\subset \bm{24_G}$.  The most stringent experimental bound on the proton lifetime comes from 
the gauge mediated proton decay mode: $p\to \pi^0e^+$ and the corresponding decay width is given by \cite{DeRujula:1980qc, FileviezPerez:2004hn, Nath:2006ut}:
\begin{align}\label{gamma}
\Gamma(p\to \pi^0e^+)= \frac{\pi m_p A^2}{2\alpha^{-2}_{GUT}M^4_{GUT}}   |\langle \pi^0|(ud)_R u_L|p\rangle|^2 \left(  |c(e^c,d)|^2+|c(e,d^c)|^2  \right),
\end{align}

\noindent
here, $m_p$ is the proton mass, the running factor of the relevant operators give $A\approx 1.8$  and, 
\begin{align}
&c(e_i,d^c_j)=V^{11}_1V^{ji}_3,\;\;c(e^c_i,d_j)=V^{11}_1V^{ij}_2+(V_1V_{UD})^{1j}(V_2V^{\dagger}_{UD})^{i1},\label{c}
\\&
V_1=U^T_cU^{\dagger},\;\;
V_2=E^T_cD^{\dagger},\;\;
V_3=D^T_cE^{\dagger},\;\;
V_{UD}=U D^{\dagger}.\label{mixing-B}
\end{align}

All the $d=6$ proton decay operators including Eq. \eqref{gamma} conserve $B-L$, as a result a nucleon can decay into a meson and an anti-lepton.
Operators that contribute to proton decay are in general model dependent. For example, in supersymmetric (SUSY) theories the most dominating contributions of  proton decay originate from  $d=5$ operators. Determination of proton decay in such cases require the knowledge of SUSY spectrum, the details of the Higgs potential and the fermion masses.  However, non-SUSY models are more predictive in this sense, because the aforementioned gauge mediated $d=6$ operators   mainly depend on the fermion mixings.  There can be additional contributions to the proton decay originating from Higgs mediated $d=6$ operators in non-SUSY models that are highly model dependent and less predictive, since apriori  the couplings entering in the scalar potential are not known.  This is why we only discuss the gauge mediated $d=6$ proton decay operator of Eq. \eqref{gamma} as they have the least model dependence.  The $c$-coefficients given in Eq. \eqref{c} depend on the detail of the flavor structure. Here to estimate the proton lifetime we take the most conservative scenario, $c(e^c,d)=2$ and $c(e,d^c)=1$ for the $p\to \pi^0e^+$ channel. This is a very good assumption since the leading entries in the mixing matrices given in Eq. \eqref{mixing-B}  that participate in the computation of $p\to\pi^0e^+$ decay are expected to have the similar structure as that of the CKM matrix  which to a very good approximation is given by  $V_{CKM}\approx \mathds{1}$. Deviation from this will only increase the proton life time and requires cancellations utilizing fine-tuned Yukawa couplings (see for example Ref. \cite{Dorsner:2004xa}) that we do not consider here. Consequently, the proton lifetime is rather  very sensitive to the unification scale $M_{GUT}$ and the associated 
unified coupling constant $\alpha_{GUT}$.    The relevant nuclear matrix element needed in Eq. \eqref{gamma} is taken from  Ref. \cite{Aoki:2017puj}:  $|\langle \pi^0|(ud)_R u_L|p\rangle|=-0.118$.  In the gauge coupling unification analysis presented below, we will demonstrate  few different scenarios and estimate the corresponding proton lifetime using Eq. \eqref{gamma}.
The  
current experimental upper bound on the proton lifetime is 
$\tau_p(p\to \pi^0e^+)> 1.6\times 10^{34}$ years \cite{Miura:2016krn} and
the  Hyper-Kamiokande experiment  after 10 years of exposure can make a $3\sigma$ discovery of $p\to \pi^0e^+$ process up to  $6.3\times 10^{34}$ years \cite{Abe:2018uyc}.

However as aforementioned, the gauge mediated processes are not the only source for the protons to decay, some of the scalar leptoquarks present in the unified theories can also lead to proton decay. 
 Note that light scalars must be present  for gauge coupling unification to realize, recall that coupling unification does not happen in the Georgi-Glashow model.   
The scalars that mediate proton decay in our theory are $\phi_2 (3,1,-\frac{1}{3}) \subset \bm{5_H}$,  $\Sigma_2 (3,1,-\frac{1}{3}), \Sigma_3 (\overline{3},1,\frac{4}{3}), \Sigma_5 (3,3,-\frac{1}{3}) \subset \bm{45_H}$ and $\chi_2 (3,1,-\frac{1}{3}) \subset \bm{50_H}$.   To suppress proton decay one would expect these fields to have masses of the order of the GUT scale. For our analysis we assume that these fields are sufficiently heavy so that the corresponding dangerous proton decay operators are suppressed. If any of these fields is assumed to be much lighter than the GUT scale, the associated Yukawa couplings need to be somewhat suppressed to avoid dangerous proton decay.   

\begin{figure}[th!]
\centering
\includegraphics[scale=0.4]{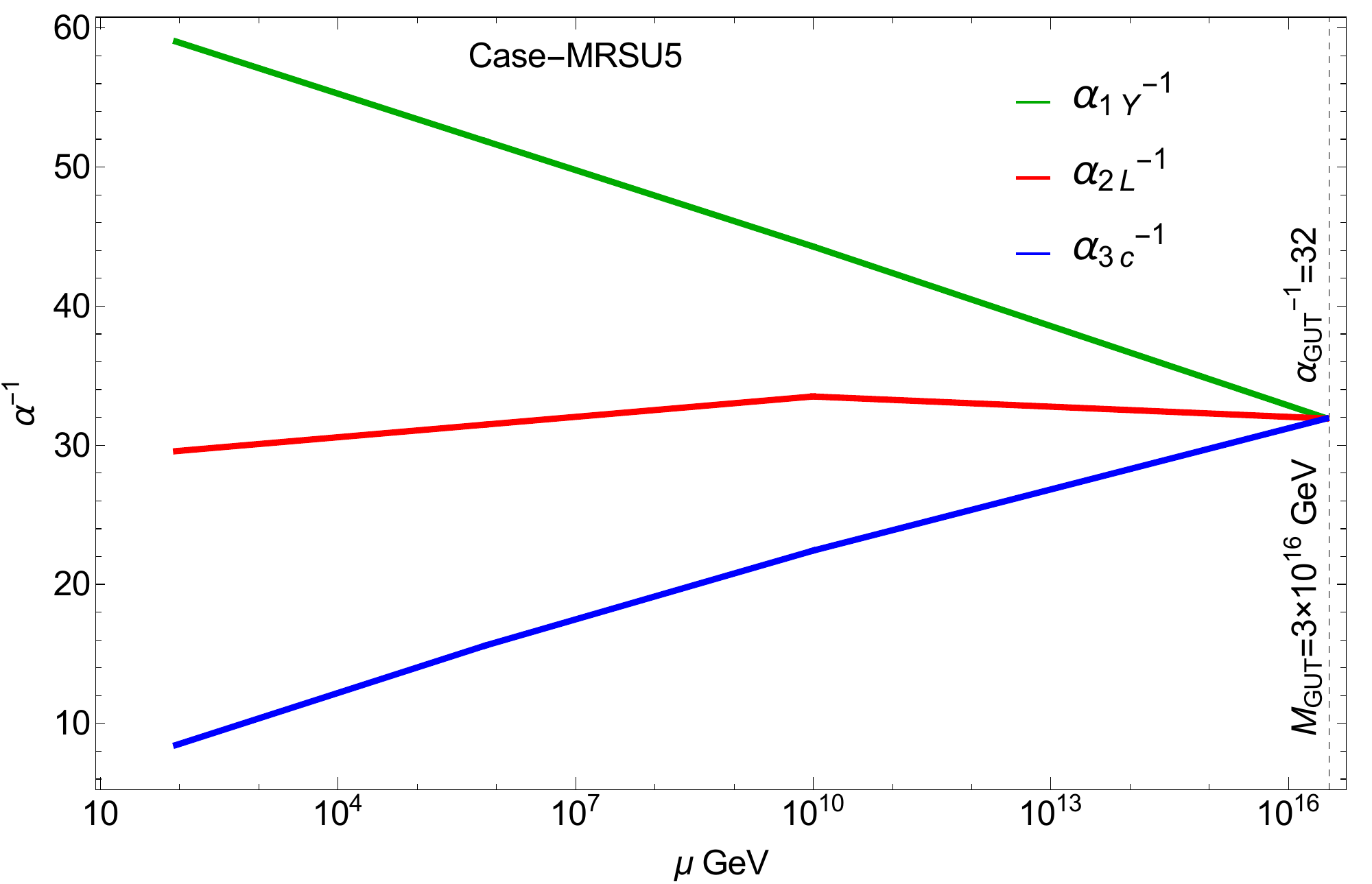}
\caption{Here we present the plot of the 
gauge coupling unification in the MRSU5 model as discussed in the text. 
}\label{unification-MRSU5}
\end{figure}

For the purpose of comparison, at first we discuss the coupling unification scenario within the minimal renormalizable model.  Note that to achieve high GUT scale value, one needs to keep scalars light that provide negative contribution to the $B_{12}$. In the MRSU5 model, other than the SM like doublets, such negative contribution is provided by the $\Phi_2, \Sigma_5, \Sigma_7$  multiplets, see table \ref{Bij}.  Among these three, $\Sigma_5$ mediate proton decay, on the other hand, $\Phi_2$ and  $\Sigma_7$ do not  and can be very light. However,  keeping $\Sigma_5$ at the GUT scale and the other two fields  light fails unification test, so $\Sigma_5$ must be light as well within this scenario. To avoid proton decay bounds, this multiplet needs to be heavier than about $10^{10}$ GeV by assuming natural values of the Yukawa couplings \cite{Dorsner:2006dj}, however, for smaller values of the Yukawa couplings, this multiplet can be kept at lower scale.    In this minimal scenario, by fixing $m_{\Sigma_5}=10^{10}$ GeV and $m_{\phi_1}= m_{\Sigma_1}= m_{\Sigma_7}= m_{\Phi_2}= M_Z$ we find, to achieve unification at the one-loop, one needs $m_{\Phi_3}=7.28\times 10^{5}$ GeV and the corresponding unification scale is $3.02\times 10^{16}$ GeV which agrees with \cite{Dorsner:2006dj}. In fig. \ref{unification-MRSU5}, we present the corresponding plot of the gauge coupling unification in this model. 

\FloatBarrier

However, this simple realization of the gauge coupling unification does not remain valid in our proposed model. 
The required scalar multiplets $\eta_1 (1,2,-\frac{3}{2})\subset \bm{40_H}$ and  $\chi_1 (1,1,-2)\subset \bm{50_H}$ running in the loop to generate neutrino mass are expected to reside in scales much smaller than the GUT scale.   The presence of these additional light scalar multiplets completely ruins the successful coupling unification of the MRSU5 model as shown above, hence, threshold corrections from other   scalar fields need to be taken into account to restore the gauge coupling unification.
 We find that in this set-up it gets difficult to achieve very high scale gauge coupling unification  without taking into account threshold correction from quite a few scalar multiplets living in between the EW and the GUT scales.

%%%%%%%%%%%%%%%%%%%%%%%%%%%%%%%%%%%%%%%%%%%%%%%
%%%%%%%%%%%%%%%%%%%%%%%%%%%%%%%%%%%%%%%%%%%%%%%
%%%%%%%%%%%%%%%%%%%%%%%%%%%%%%%%%%%%%%%%%%%%%%%
\FloatBarrier
\begin{table}[t!]
\centering
\footnotesize
\resizebox{1\textwidth}{!}{
\begin{tabular}{|c|c|c|c|c|c|}
\hline
Case &Multiplets& Mass (in GeV) 
& $M_{\rm GUT}$ (GeV) 
& $\alpha^{-1}_{\rm GUT}$ 
& $\tau_p(p\to \pi^0e^+)$ in years  \\ \hline\hline

A& 
\pbox{10cm}{
\vspace{2pt}
$\Phi_{3}, \chi_1, \eta_{3,4,5}$\\
$\eta_{1,2,6}$\\
$\Sigma_5$
\vspace{3pt}
}
& 
\pbox{10cm}{
\vspace{2pt}
$10^3$\\
$3.90\times 10^6$\\
$1.58\times 10^{11}$
\vspace{3pt}
}
&$6.2\times 10^{15}$&$16.2$ &$2.43\times 10^{34}$  \\ \hline\hline

B&
\pbox{10cm}{
\vspace{2pt}
$\Phi_{3}, \eta_{1,3,4,5}$\\
$\chi_1, \eta_{2,6}$\\
$\Sigma_5$
\vspace{3pt}
}
&
\pbox{10cm}{
\vspace{2pt}
$10^3$\\
$2.33\times 10^6$\\
$1.58\times 10^{11}$
\vspace{3pt}
}
&$5.37\times 10^{15}$&$15.9$&$1.32\times 10^{34}$
\\ \hline\hline

C&
\pbox{10cm}{
\vspace{2pt}
$\Phi_{3}, \chi_{1}$\\
$\Sigma_5$\\
$\eta_1$
\vspace{3pt}
}
&
\pbox{10cm}{
\vspace{2pt}
$10^3$\\
$2\times 10^{7}$\\
$2.9\times 10^{14}$
\vspace{3pt}
}
&$1.88\times 10^{15}$&$30.3$&$7.2\times 10^{32}$
\\ \hline\hline

D&
\pbox{10cm}{
\vspace{2pt}
$\Phi_{3}, \chi_1, \eta_{3,4,5}$\\
$\eta_1$\\
$\eta_6$\\
$\eta_2$\\
$\Sigma_5$
\vspace{3pt}
}
&
\pbox{10cm}{
\vspace{2pt}
$10^3$\\
$4.14\times 10^4$\\
$7.68\times 10^6$\\
$2.72\times 10^7$\\
$1.58\times 10^{11}$
\vspace{3pt}
}
&$4.25\times 10^{15}$&$16.33$&$5.46\times 10^{33}$
\\ \hline\hline

E&
\pbox{10cm}{
\vspace{2pt}
$\chi_1, \eta_{1}$\\
$\Phi_3, \eta_5$\\
$\eta_{2,6}$\\
$\eta_3$\\
$\Sigma_5$
\vspace{3pt}
}
&
\pbox{10cm}{
\vspace{2pt}
$10^3$\\
$3.5\times 10^3$\\
$7.5\times 10^4$\\
$4.8\times 10^7$\\
$1.58\times 10^{9}$
\vspace{3pt}
}
&$2.2\times 10^{17}$&$10.9$&$1.74\times 10^{40}$
\\ \hline

\end{tabular}
}
\caption{ 
Successful 
gauge coupling unification  scenarios within our framework. For all these five cases, the two iso-spin doublet masses are taken to be  $m_{\phi_1, \Sigma_1}= v_{EW}$ and the masses of the $\Phi_2$ and $\Sigma_7$ multiplets are fixed at $m_{\Phi_2}=1$ TeV and  $m_{\Sigma_7}=3.5$ TeV.  It should be pointed out that for the cases C and D where the proton life times are estimated to be somewhat smaller than the current upper bound $\tau_p(p\to \pi^0e^+)>1.6\times 10^{34}$ years,  small threshold corrections near the GUT scale can make these scenarios viable. The corresponding gauge coupling unification plots are presented in Fig. \ref{unification}.   
}
\label{examples}
\end{table}
%%%%%%%%%%%%%%%%%%%%%%%%%%%%%%%%%%%%%%%%%%%%%%%
%%%%%%%%%%%%%%%%%%%%%%%%%%%%%%%%%%%%%%%%%%%%%%%
%%%%%%%%%%%%%%%%%%%%%%%%%%%%%%%%%%%%%%%%%%%%%%%

With five  different scenarios  (we label them as A, B, C, D, E) we demonstrate how gauge coupling unification can be restored in our model.  
   For this analysis, we fix the masses of the two SM like iso-spin doublets as $m_{\phi_1, \Sigma_1}= v_{EW}$. We also fix $m_{\Sigma_7}=3.5$ TeV, since from the collider bounds it is required that $m_{\Sigma_7} > 3.1$ TeV provided that the Yukawa couplings take natural values \cite{Khachatryan:2015dcf}.  We also take $m_{\Phi_{2}}= 1$ TeV and furthermore,  
for the cases A, C, D: $m_{\chi_1}=1$ TeV,  
for case B: $m_{\eta_1}=1$  TeV
and 
for case E: $m_{\eta_1}=m_{\chi_1}=1$ TeV are assumed.    
With these assumptions and using the mass relations derived in Sec. \ref{SEC-04}, the results are presented in the table \ref{examples} with the corresponding unified value of the gauge coupling constant, the scale of the unification and the estimation of the associated proton lifetime for each scenario.  It should be pointed out that  for cases C and D, even though the proton life time are estimated to be somewhat below the current upper bound $\tau_p(p\to \pi^0e^+)>1.6\times 10^{34}$ years, small threshold corrections near to the GUT scale can make these scenario viable. Even though these choices are not unique, but clearly demonstrate how successful gauge coupling unification consistent  with proton decay bounds can be achieved within our set-up. 
Due to the presence of the light scalars that play role in neutrino mass generation, unification scale cannot be made arbitrarily large and the proton decay rate is expected to be within the observable range. 
 Though no firm prediction can be made about the proton decay within this framework, but for most of the examples provided here,  the proton decay rate is very close to the current experimental bound and has the potential to be tested in near future.

\begin{figure}[th!]
\centering
\includegraphics[scale=0.4]{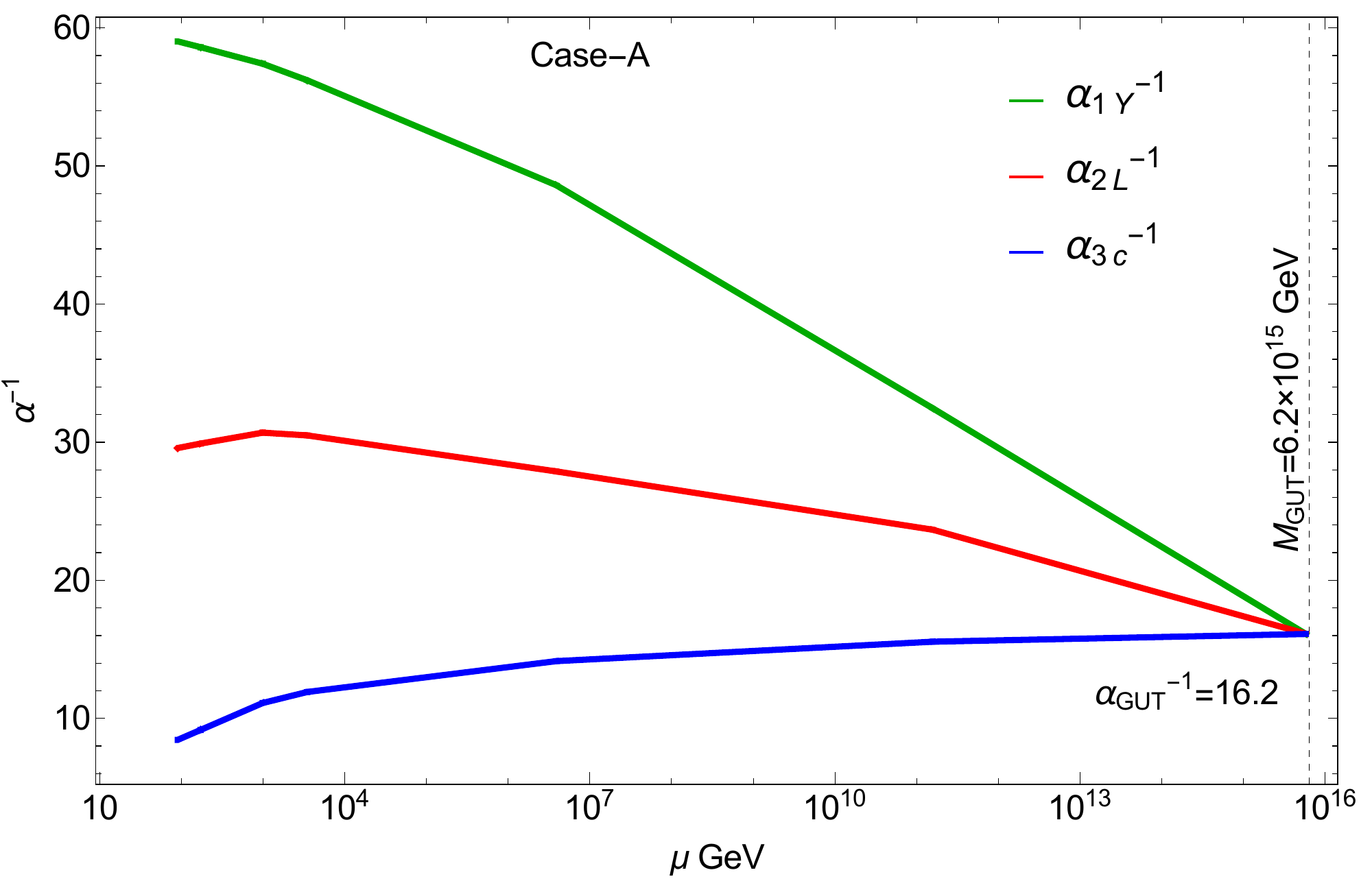}
\includegraphics[scale=0.4]{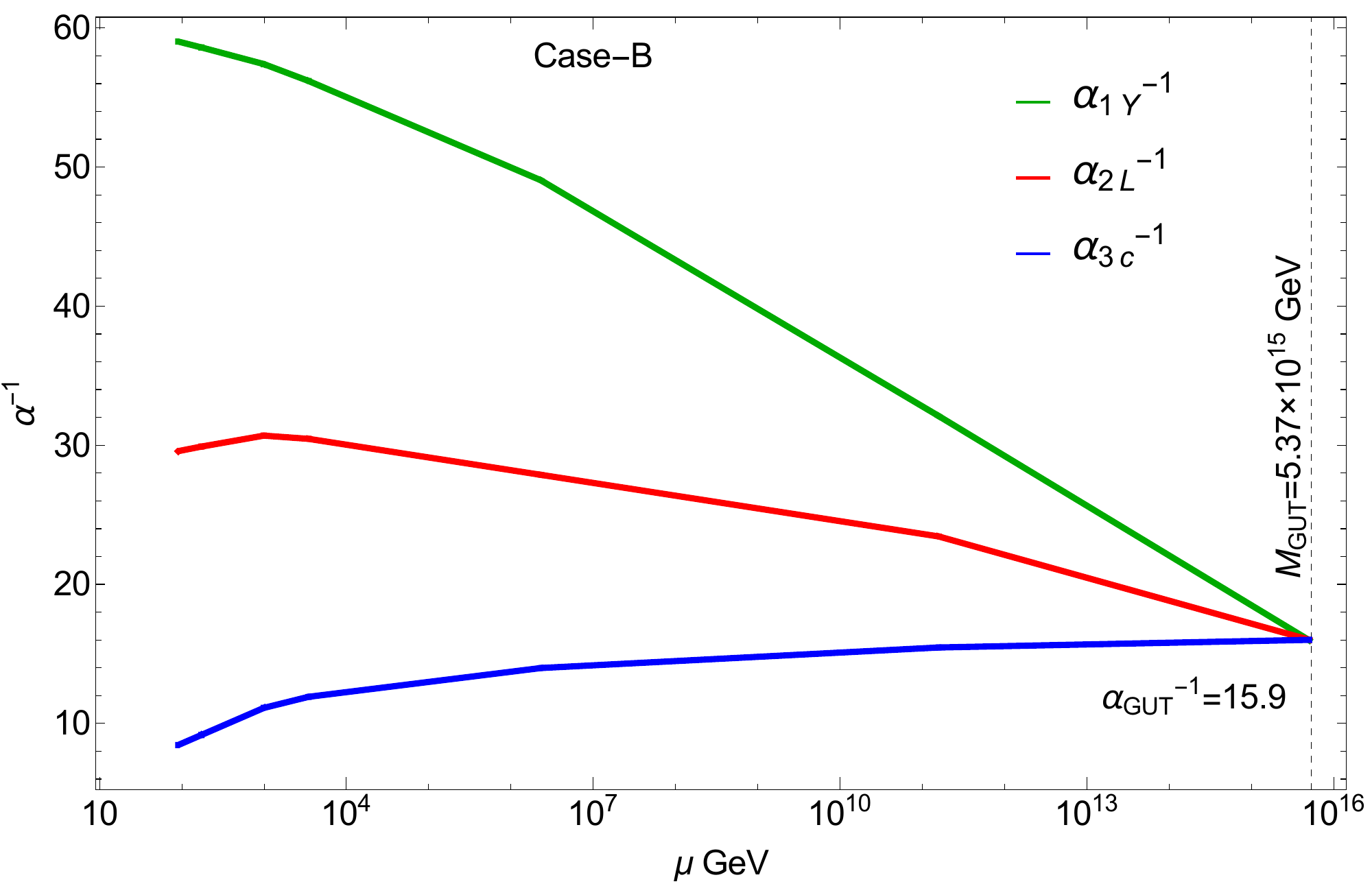}
\includegraphics[scale=0.4]{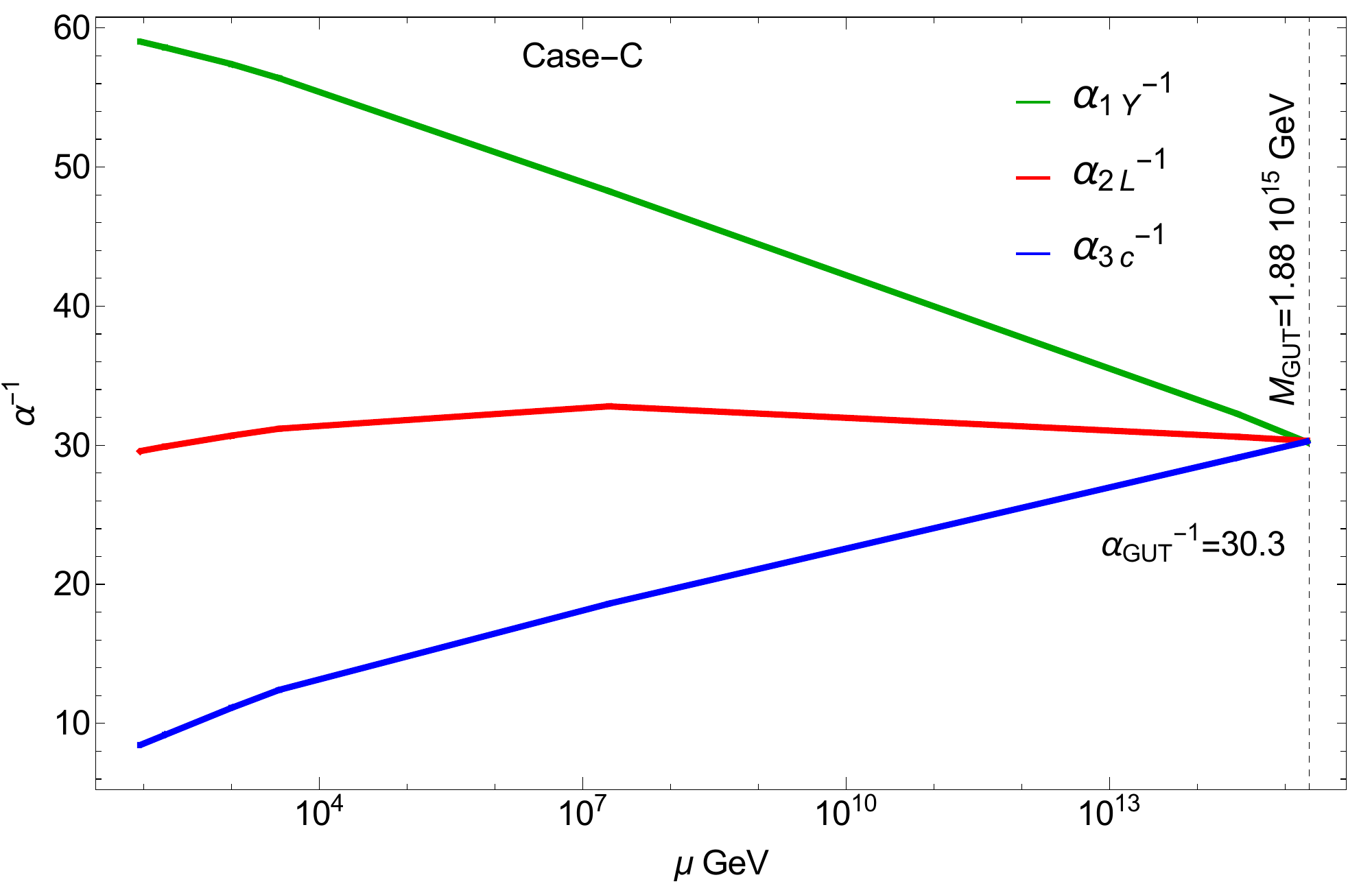}
\includegraphics[scale=0.4]{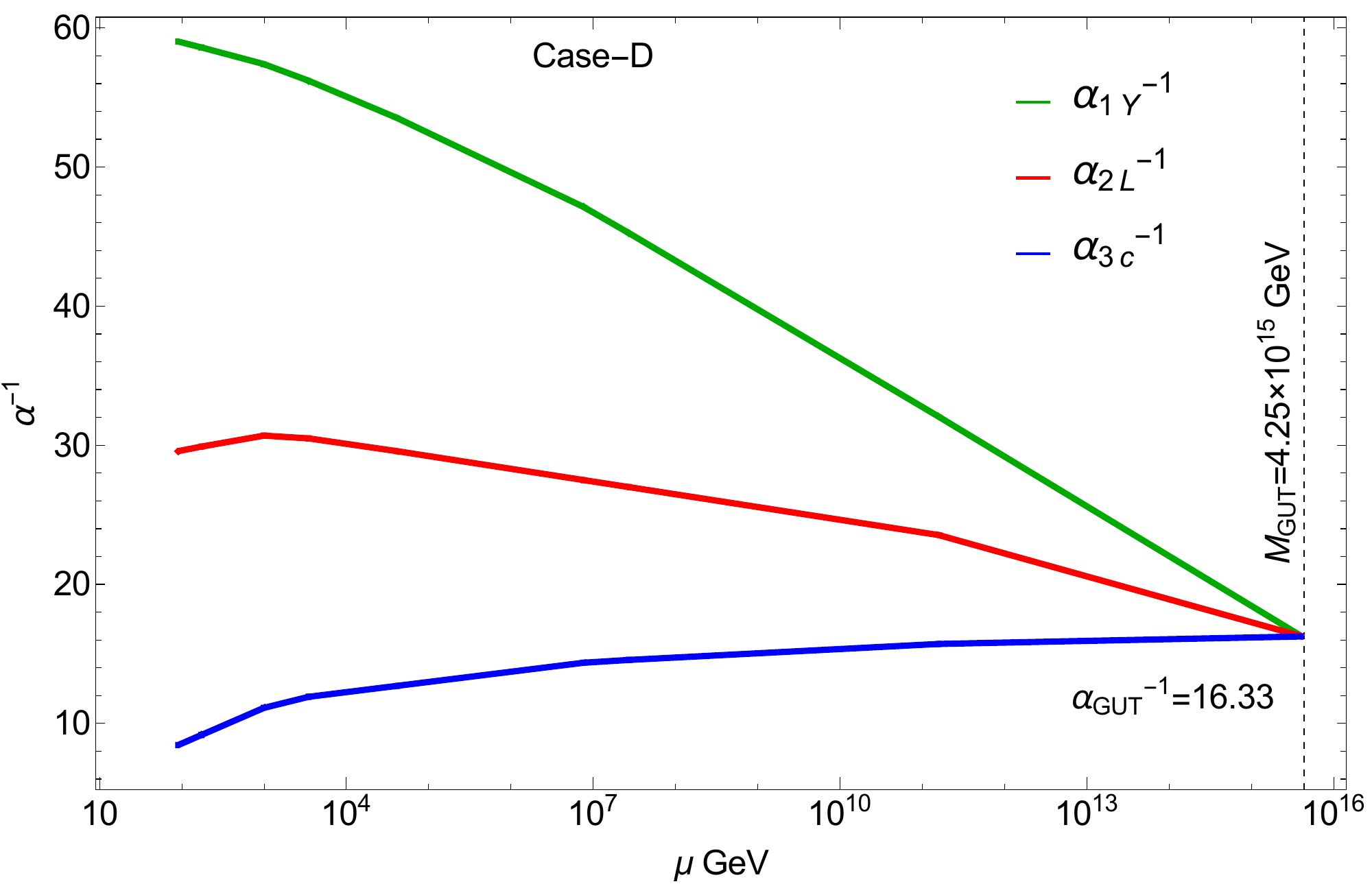}
\includegraphics[scale=0.4]{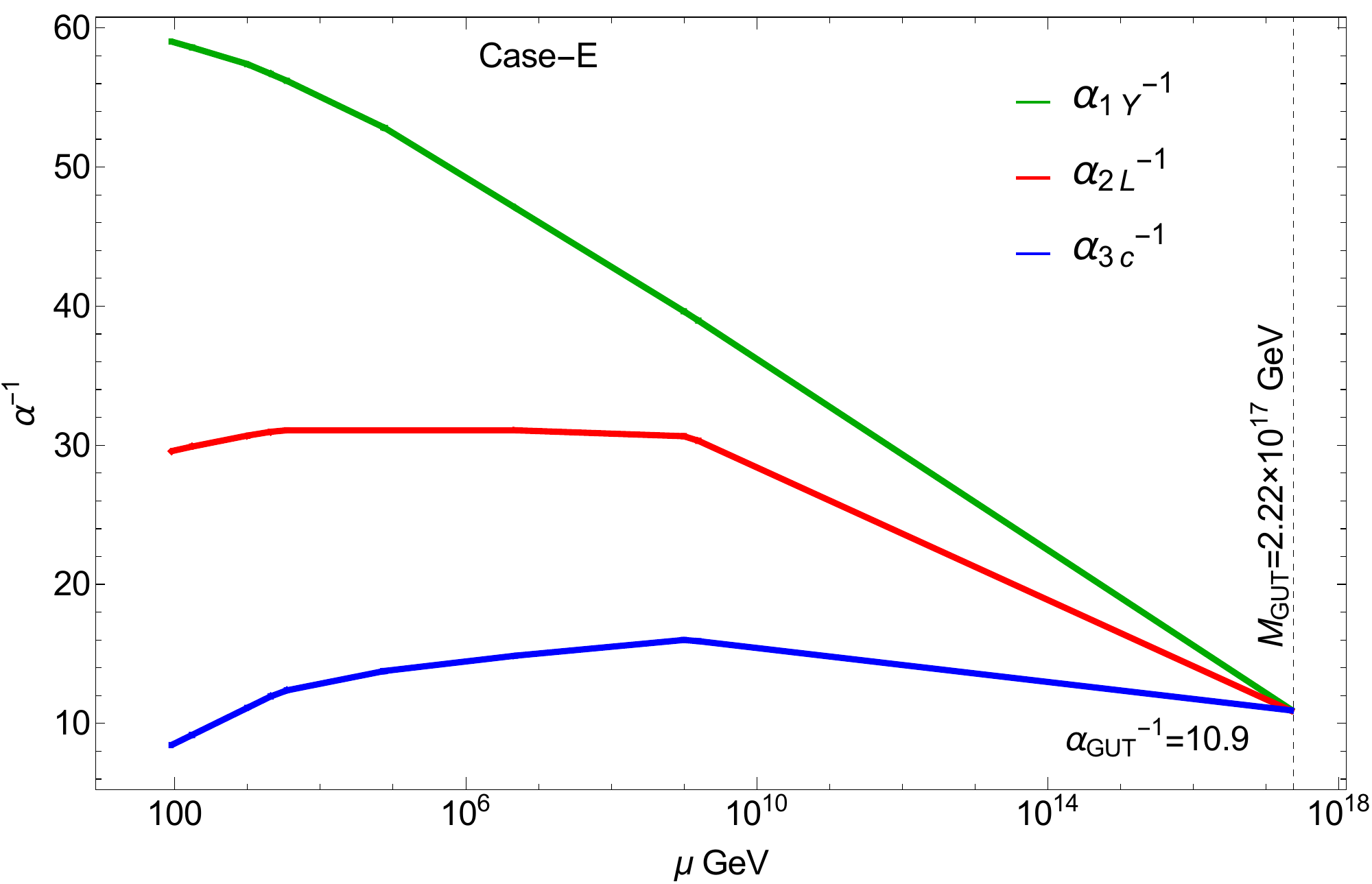}
\caption{Here we present the plots of the 
gauge coupling unification scenarios within our framework that are  summarized in Table  \ref{examples}. 
}\label{unification}
\end{figure}

For completeness, in Fig. \ref{unification} we present the plots of the gauge coupling unification for the aforementioned five scenarios that are  summarized in Table \ref{examples}. 
 As already pointed out, successful  gauge coupling unification in our set-up requires  more number of light scalar multiplets  compared to the minimal model (MRSU5). In coherence with  the MRSU5 case, to achieve unification we keep the fields $\Phi_{2,3}$, $\Sigma_7$  around the TeV range and the multiplet $\Sigma_5$ not too far from $10^{10}$ GeV.  However, since $\chi_1$ and  $\eta_1$ fields are expected to be much smaller than the GUT scale in our framework, to compensate for their effects on the running of the coupling constants, few more additional fields must reside in between the EW and the GUT scales.
 For most of the cases (A, B, C, D) considered here, even with quite a few new multiplets (different set of multiplets for different cases) living at the low energies, 
 the scales of unification are found to be an order of magnitude less compared to the minimal set-up. We also demonstrate a scenario (case E) where unification scale can be achieved which  is an order higher compared to the case of MRSU5.

\FloatBarrier

%%%%%%%%%%%%%%%%%%%%%%%%%%%%%%%%%%%%%%%%%%%%%%%
%%%%%%%%%%%%%%%%%%%%%%%%%%%%%%%%%%%%%%%%%%%%%%%
\section{Conclusions}\label{SEC-08}
Grand unification based on the $\bm{SU(5)}$ gauge group is one of the leading candidates for the ultraviolet completion of the SM. 
The minimal $\bm{SU(5)}$ GUT has many attractive features, however fails to incorporate neutrino mass. 
In this work, we have proposed a renormalizable $\bm{SU(5)}$ GUT 
where neutrino mass originates at the two-loop level.
By detail analysis we have shown that this proposed model is the only \textit{true two-loop} model of neutrino mass generation based on $\bm{SU(5)}$ GUT in the existing literature.   
 This realization requires two  Higgs representations beyond the minimal renormalizable model  and no additional fermion beyond the SM is introduced.  Within this set-up, 
in addition to correctly reproducing the charged fermion and neutrino mass spectrum, successful gauge coupling unification can be realized  
  while simultaneously satisfying the proton decay bounds. 
  It is shown that the neutrino masses are not completely independent but are correlated with the charged fermion masses. 
By constructing the relevant scalar potential, the Higgs mass spectrum is computed and few examples of gauge coupling unification are demonstrated.   
Proton decay rate is expected to be within the observable range in our framework.       The Higgs representations  required for generating realistic fermions masses contain leptoquarks that can accommodate the recent B-physics anomalies \footnote{
Recently, the B-physics anomalies have gained a lot of attention in the high energy physics community, particularly the lepton flavor universality  ratios 
$R_{K^{(\ast)}}$ and $R_{D^{(\ast)}}$. The deviations of the measurements on $R_{K^{(\ast)}}$ \cite{Aaij:2014ora,Aaij:2017vbb} and $R_{D^{(\ast)}}$ \cite{Lees:2013uzd,Hirose:2016wfn,Aaij:2015yra} from the SM are within $2-3 \sigma$ confidence level. 
In Ref. \cite{Becirevic:2018afm} it is pointed out  that a TeV scale LQ with quantum number  $(\overline{3},2,-\frac{7}{6})$ originating from $\bm{45_H}$  and $\bm{50_H}$  and a second LQ with quantum number $(3,3,-\frac{1}{3})$ residing at the sub-TeV scale  originating from $\bm{45_H}$ can simultaneously explain both these anomalies.  
Since our set-up has both the $\bm{45_H}$ and $\bm{50_H}$ representations, we expect that the proposed model in this work along with explaining the origin of neutrino mass can also incorporate the observed  B-anomalies.
Whereas establishing such a link among these seemingly different phenomena in a unified framework is interesting,   however  is  beyond the scope of the present work. 
For a detailed analysis on how to 
 accommodate B-physics anomalies in the context of the framework  discussed in this article,  we refer the readers to this work \cite{Becirevic:2018afm}.}.

%%%%%%%%%%%%%%%%%%%%%%%%%%%%%%%%%%%%%%%%%%%%%
%%%%%%%%%%%%%%%%%%%%%%%%%%%%%%%%%%%%%%%%%%%%%
\section*{Acknowledgments}
The author would like to thank  Dr.	Ilja Dorsner for discussion. 

%%%%%%%%%%%%%%%%%%%%%%%%%%%%%%%%%%%%%%%%%%%%%
%%%%%%%%%%%%%%%%%%%%%%%%%%%%%%%%%%%%%%%%%%%%%
%\newpage
\FloatBarrier

\end{document}